\documentclass{aa}

\usepackage{graphicx}
\usepackage{txfonts}
\usepackage{natbib,twoopt}
\usepackage[breaklinks=true]{hyperref} 
\bibpunct{(}{)}{;}{a}{}{,}             
\makeatletter
  \newcommandtwoopt{\citeads}[3][][]{\href{http://adsabs.harvard.edu/abs/#3}%
    {\def\hyper@linkstart##1##2{}%
     \let\hyper@linkend\@empty\citealp[#1][#2]{#3}}}
  \newcommandtwoopt{\citepads}[3][][]{\href{http://adsabs.harvard.edu/abs/#3}%
    {\def\hyper@linkstart##1##2{}%
     \let\hyper@linkend\@empty\citep[#1][#2]{#3}}}
  \newcommandtwoopt{\citetads}[3][][]{\href{http://adsabs.harvard.edu/abs/#3}%
    {\def\hyper@linkstart##1##2{}%
     \let\hyper@linkend\@empty\citet[#1][#2]{#3}}}
  \newcommandtwoopt{\citeyearads}[3][][]%
    {\href{http://adsabs.harvard.edu/abs/#3}
    {\def\hyper@linkstart##1##2{}%
     \let\hyper@linkend\@empty\citeyear[#1][#2]{#3}}}
\makeatother
\usepackage{amstext}

\begin{document}

   \title{New ultracool dwarf neighbours within 20\,pc from {\it Gaia} DR2}


\titlerunning{New UCD neighbours within 20\,pc from {\it Gaia} DR2}

   \author{R.-D. Scholz\inst{1}
          }

   \institute{Leibniz-Institut f\"ur Astrophysik Potsdam,
              An der Sternwarte 16, D--14482 Potsdam, Germany\\
              \email{rdscholz@aip.de}
             }

   \date{Received December 20, 2019; accepted March 23, 2020}


  \abstract
   {}
   {The {\it Gaia} data release 2 (DR2) contains $>$6000 objects with
parallaxes $(Plx+3\times{e\_Plx})>50$\,mas placing them 
within 20\,pc from the 
Sun. Since the expected numbers 
extrapolating the well-known 10\,pc census are much lower, 
nearby {\it Gaia} stars need a quality assessment.
The 20\,pc sample of white dwarfs (WDs) had already been verified
and completed with {\it Gaia} DR2. We aimed 
to check and complete 
the 20\,pc sample of ultracool dwarfs (UCDs) with spectral types $\gtrsim$M7
and given {\it Gaia} DR2 parallaxes.}
   {Dividing the {\it Gaia} DR2 20\,pc sample
into subsamples of various astrometric and 
photometric quality,
we studied their distribution on the sky, in the 
$M_G$ vs. $G$$-$$RP$ colour-magnitude diagram (CMD), and
as a function of  $G$ magnitude and total proper motion. 
After excluding 139 
known WDs 
and 263 known UCDs from the CMD, we checked all
remaining $\approx$3500 candidates with $M_G>14$\,mag 
(used to define UCDs in this study)
for the correctness
of their {\it Gaia} DR2 proper motions 
via 
visual inspection of finder charts, 
comparison with proper motion catalogues, and 
comparison with our own proper motion measurements.
For 
confirmed UCD candidates 
we estimated spectral types
photometrically using {\it Gaia} and near-infrared absolute magnitudes
and colours.}
   {We failed to confirm new WDs, but found 50 new UCD
candidates not mentioned in three previous studies using 
{\it Gaia} DR2. They have relatively small proper motions and 
tangential velocities and are concentrated towards the Galactic plane. 
Half of them have spectral types in SIMBAD and/or 
previous non-{\it Gaia} distance estimates that placed them
already within 20\,pc. 
For 20 of the 50 objects, we estimated
photometric spectral types of M6-M6.5, slightly below the classical 
UCD spectral type limit. However,
seven L4.5-L6.5, four L0-L1, five M8.5-M9.5, and three M7-M8 dwarfs
can be considered as completely new UCDs discoveries within 20\,pc
based on {\it Gaia} DR2. Four M6.5 and two L4.5 dwarfs have high
membership probabilities (64\%-99\%) in the ARGUS, AB Doradus, or
Carina Near young moving groups.} 
   {}

   \keywords{
Parallaxes --
Proper motions --
brown dwarfs --
Stars: distances --
Hertzsprung-Russell and C-M diagrams --
solar neighbourhood
               }

   \maketitle


\section{Introduction}
\label{Sect_intr}

Objects of spectral types M7 and later were first called ultracool 
dwarfs (UCDs) by \citetads{1997AJ....113.1421K}. This definition 
by spectral type $\ge$M7 is still being used for classical UCDs.
With cooler and cooler UCDs
discovered with time, three additional spectral classes, L, T, and Y had to
be created \citepads[see][for a historical review and brief discussion of
the changing physical conditions and resulting spectral morphology
of UCDs]{2014ASSL..401..113C}. From individual dynamical masses of M7-T5
UCDs in the
solar neighbourhood, \citetads{2017ApJS..231...15D} found that a spectral
type of L4 and mass of 70 Jupiter masses roughly correspond to the
hydrogen-burning limit. However, a spectral type of L4 can not be
considered as a boundary between stellar and substellar objects
in the solar neighbourhood,
since their ages are different (and difficult to
determine) and brown dwarfs change their spectral types as they cool
down.

Despite many long-term efforts 
the solar neighbourhood is still
not fully explored, and even at very short distances ($<$10\,pc)
from the 
Sun
new discoveries of a few white dwarfs (WDs), 
up to $\approx$10\% more red dwarf stars,
and probably at least 20\% more brown dwarfs can be expected
\citepads{2018AJ....155..265H,2019ApJS..240...19K}.
The knowledge about our nearest stellar and substellar neighbours
is important for many aspects of astronomical research.
This includes the estimated outcome of star formation 
\citepads[e.g.][]{2013AN....334...26K,2019ApJS..240...19K},
the local density and velocity distribution
and the role of (sub)stellar encounters 
\citepads[e.g.][]{2014A&A...561A.113S,2015ApJ...800L..17M,
2015A&A...575A..35B,2017MNRAS.464.2290M} 
and interstellar comets 
\citepads[e.g.][]{2018MNRAS.476L...1D,2018MNRAS.479L..17P,2019arXiv191102473H}.
The search 
for extrasolar planets and investigation of
planetary systems around low-mass stars 
\citepads[e.g.][]{2016Natur.536..437A,2017Natur.542..456G,2018Natur.563..365R}
and brown dwarfs
\citepads{2019sptz.prop14257M,2019hst..prop15884B}
can also be carried out very effectively in our close neighbourhood.
Some of the brown dwarfs crossing the solar neighborhood may be
relatively young ($\approx$10-100\,Myr). They can still be relatively bright
and may have not yet cooled down to T or Y types,
and an opportunity for age determinations arises from their kinematic 
membership 
in nearby young moving groups
\citepads[YMGs;][]{2016ApJ...821..120A,2017ApJ...841L...1G,2019AJ....157..247R}.

According to the statistics of the REsearch Consortium On Nearby Stars
(RECONS, see www.recons.org), in the 
time from 2000 to 2018, the total number of stars and brown dwarfs
within 10\,pc from the Sun rose from 291 to 428.
This increase was achieved mainly 
thanks to 85 new M dwarfs
and 49 new LTY dwarfs, but also three new WDs
uncovered in the immediate solar neighbourhood.
In 2000, only 198 M dwarfs, one of the LTY dwarfs, and 18 WDs
were known.
The RECONS 10\,pc census for 2018 was compiled before 
the second data release of {\it Gaia}
\citepads[DR2;][]{2018A&A...616A...1G}.
Incorporating results from
{\it Gaia} DR2, \citetads{2019AAS...23325932H} provided
a slightly smaller total number of 418 individual objects
and mentioned that 15\% of the previously known systems are
missing in {\it Gaia} DR2, whereas only 8 systems were
added by {\it Gaia} to the 10\,pc sample so far. We are aware of only
one published new system within 10\,pc, consisting of a WD/M dwarf pair
discovered already based on {\it Gaia} DR1 data 
\citepads{2018A&A...613A..26S}, and assume the other 
seven unpublished 
{\it Gaia}-based discoveries to consist of early- and mid-M dwarfs. 
 
For {\it Gaia}, observing in the optical, many of the
cooler brown dwarfs represent difficult targets at the limiting 
magnitude of the survey. Assuming a limiting magnitude $G<20.7$\,mag,
\citetads{2017MNRAS.469..401S} predicted distance limits for
L0-T9 dwarfs 
observable by {\it Gaia}. 
According to their Table~1, all $\lesssim$L6 dwarfs within 23\,pc
should be detectable by {\it Gaia}. With later L types,
the {\it Gaia} distance limit
continuously falls to 19\,pc at L7, 15\,pc at L8, and 12\,pc at L9.
However, it rises again to 14\,pc for all T0-T4 dwarfs, before
dropping to 12\,pc at T5, 10\,pc at T6, ..., and finally 2\,pc at T9.

Compared to the above described RECONS 10\,pc census, 
the number of 1722 {\it Gaia} DR2
entries with parallaxes larger than 100\,mas appears much too high.
\citetads{2019AAS...23325932H} estimated that at least 80\% of these
data are unreliable and ''only careful vetting will reveal the real members''
of the 10\,pc sample. Here, we study the {\it Gaia} DR2 20\,pc sample
(an eight times larger volume) in Sect.~\ref{Sect_20pc} with respect to
various astrometric and photometric quality criteria (Sect.~\ref{SubSect_q}).
In contrast to the classical UCD definition by spectral type ($\ge$M7), 
we define an absolute {\it Gaia} magnitude limit of $M_G>14$\,mag
for all UCDs considered in this study and review previous studies of UCDs
and WDs with respect to the {\it Gaia} DR2 20\,pc sample 
(Sect.~\ref{SubSect_ucdwd}). Finally, we describe our search
for new UCDs and WDs ((Sect.~\ref{SubSect_search}). In our search
we tried to find 
nearby 
UCDs in particular among the faintest {\it Gaia} objects, 
which have typically unreliable data.
All detected UCDs that were not yet included in three previous
UCD-related studies involving {\it Gaia} DR2 data are considered as
''new UCDs'' in this study.
In Sect.~\ref{Sect_photclass}, we classify 
these new UCDs found in
{\it Gaia} DR2 photometrically, in Sect.~\ref{Sect_kinspat},
we briefly investigate their spatial distribution and kinematics,
and in Sect.~\ref{Sect_YMG} we study their membership in
nearby young moving groups (YMGs),
before we conclude with an outlook to future {\it Gaia} data releases
(Sect.~\ref{Sect_outlook}).

   \begin{figure}
   \centering
   \includegraphics[width=\hsize]{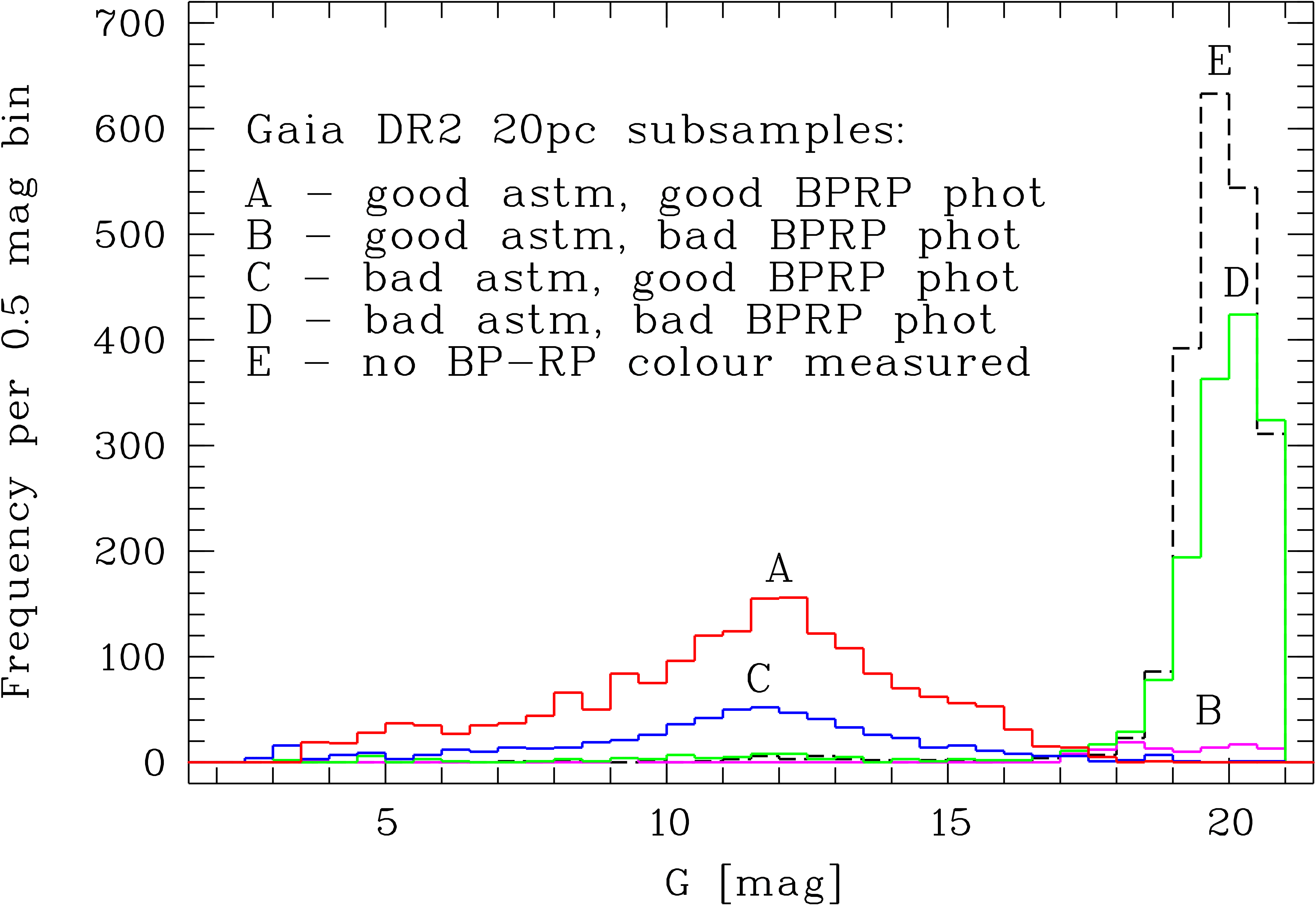}
      \caption{{\it Gaia} DR2 $G$ magnitudes of all
              objects in the 20\,pc sample: for
              different astrometric and photometric quality
              (subsamples A-D, see Sect.~\ref{SubSect_q}, shown
              in red, magenta, blue, and green) and for
              all objects without $BP$$-$$RP$ colour (subsample E,
              black dashed histogram).
              }
      \label{Fig_Ghist}
   \end{figure}

\section{The {\it Gaia} DR2 20\,pc sample}
\label{Sect_20pc}

To select 
an initial {\it Gaia} DR2 20\,pc sample we applied a
soft parallax cut using the 
parallax errors $(Plx+3\times{e\_Plx})>50$\,mas,
as it was also done by \citetads{2018MNRAS.480.3942H}
to define
their {\it Gaia} 20\,pc WD sample. This soft parallax cut yielded 
6105 objects, whereas with a sharp cut of $Plx>50$\,mas we found only
5400 objects. Both numbers are much higher than the expected number
of $8\times418\approx3350$ objects within 20\,pc, derived from the latest 
estimate for the 10\,pc sample \citepads{2019AAS...23325932H} and assuming a
constant density of objects. On the other hand, we think that {\it Gaia} DR2
may have missed a similarly large fraction of real members of the 20\,pc 
sample as found by \citetads{2019AAS...23325932H} for the 
10\,pc sample ($\approx$15\%). 

The parallax errors of all 6105
objects are smaller than 5\,mas. This means that almost all (except for 14 
objects) of them have formally better than 10\% parallax precision.
There is a strong correlation of the parallax errors with the $G$ 
magnitude. Whereas between $G=5$\,mag and $G=17$\,mag the median 
of the parallax error is always below 0.1-0.2\,mas (with outliers
reaching 1-2\,mas), it raises systematically to 0.5\,mas at the bright end
($G=3$\,mag) and 3.5\,mas at the faint end ($G=21$\,mag).
With the formally high precision of the parallaxes in the 
{\it Gaia} DR2 20\,pc sample, we can compute absolute magnitudes as
$M_G=G+5\times\log{(Plx/100)}$, where the parallax $Plx$ is in
units of mas.

Most of the objects in our {\it Gaia} DR2 20\,pc sample are rather 
faint ($G>19$\,mag), where both astrometric and photometric measurements
in {\it Gaia} DR2 are problematic (see Sect.\ref{SubSect_q}). In
addition, many 
predominantly faint objects lack colour information,
as shown by the dashed histogram in Fig.~\ref{Fig_Ghist}. 
The $BP$$-$$RP$
colour measurements are used for assessing the photometric quality,
in particular in crowded fields, e.g. in the Galactic plane
(Sect.\ref{SubSect_q}). Only 4065 out of the 6105 objects have
$BP$$-$$RP$ colours and 4102 have $G$$-$$RP$ colours 
measured in {\it Gaia} DR2.

Observational Hertzsprung-Russell or colour-magnitude 
diagrams (CMDs) 
with absolute magnitudes 
determined directly from
the parallaxes have been presented by
\citetads{2018A&A...616A..10G}, and \citetads{2018A&A...616A...2L}
demonstrated how the CMD 
is getting cleaner when astrometric and
photometric quality criteria, similar to those described in 
Sect.~\ref{SubSect_q}, are applied. 
The {\it Gaia} DR2 CMDs displaying 
absolute magnitudes $M_G$ as a function of $G$$-$$RP$ colour show a better
separation between the main sequence (MS) and the WD sequence
than CMDs with $M_G$ vs. $BP$$-$$RP$ colour \citepads{2019MNRAS.485.4423S}.
Therefore, we also use the $G$$-$$RP$ colour, if available, 
in the CMDs included in the following subsections
(Figs.~\ref{Fig_CMDall}, \ref{Fig_CMDallknown}, and
\ref{Fig_CMDallnoknown}).

\subsection{astrometric and photometric quality criteria}
\label{SubSect_q}

Following the recommendations given
in \citetads{2018A&A...616A...1G} 
and \citetads{GAIA-C3-TN-LU-LL-124-01},
we selected four subsamples with different quality criteria
from all objects in the {\it Gaia} DR2 20\,pc sample with available
$BP$$-$$RP$ colour measurements. These subsamples were defined 
by us as:
\begin{description}
\item[A] - good astrometry AND good photometry,
\item[B] - good astrometry BUT bad photometry,
\item[C] - bad astrometry BUT good photometry,
\item[D] - bad astrometry AND bad photometry.
\end{description}

We considered the astrometry as ''good'', when all of the 
following three criteria were fulfilled (otherwise ''bad''):
\begin{description}
\item[astm\_q1] - $Plx/e\_Plx>10$,
\item[astm\_q2] - $RUWE<1.4$,
\item[astm\_q3] - \textsf{visibility\_periods\_used}$>8$.
\end{description}
As the most important astrometric quality parameter in addition to the 
obviously important ratio $Plx/e\_Plx$,
\citetads{GAIA-C3-TN-LU-LL-124-01} introduced the 
renormalised unit weight error ($RUWE$), and we apply the same limit
of 1.4 that he suggested 
for well-behaved single sources, 
not affected by close binary companions or background objects
in crowded fields. 
The $RUWE$ corresponds to the formerly
used simple unit weight error $u$ \citepads{2018A&A...616A...2L}, 
after normalisation over both $G$ magnitude and $BP$$-$$RP$ colour.
The parameter \textsf{visibility\_periods\_used}
gives the number of groups of observations
during the only 22 months used for {\it Gaia} DR2. The minimum number
for all objects with parallaxes included in {\it Gaia} DR2 
was \textsf{visibility\_periods\_used}$=$6. \citetads{2018A&A...616A..10G} 
applied \textsf{visibility\_periods\_used}$>$8 in their study
of {\it Gaia} DR2 CMDs, as we do here.

   \begin{figure}
   \centering
   \includegraphics[width=73mm]{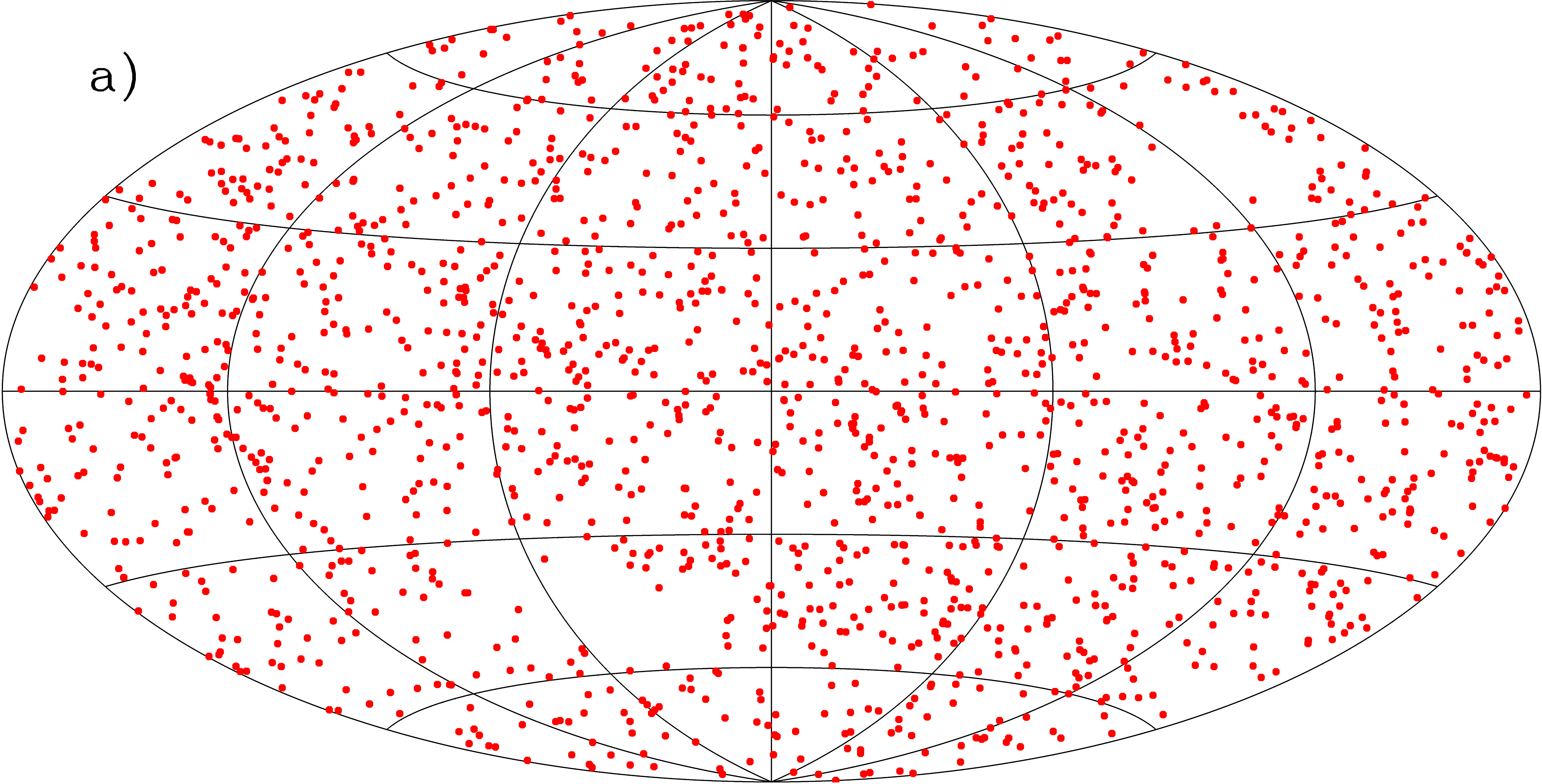}
   \includegraphics[width=73mm]{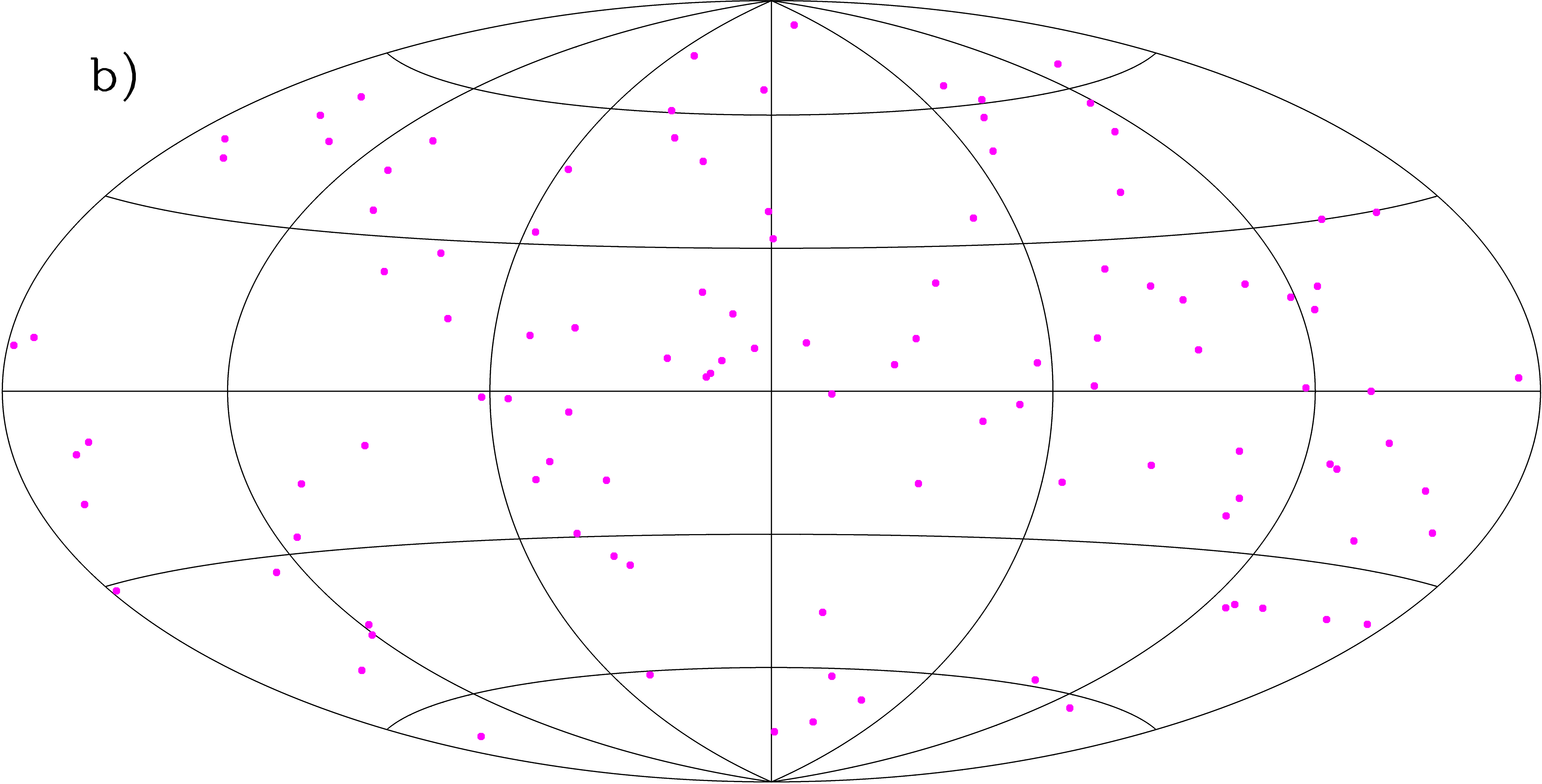}
   \includegraphics[width=73mm]{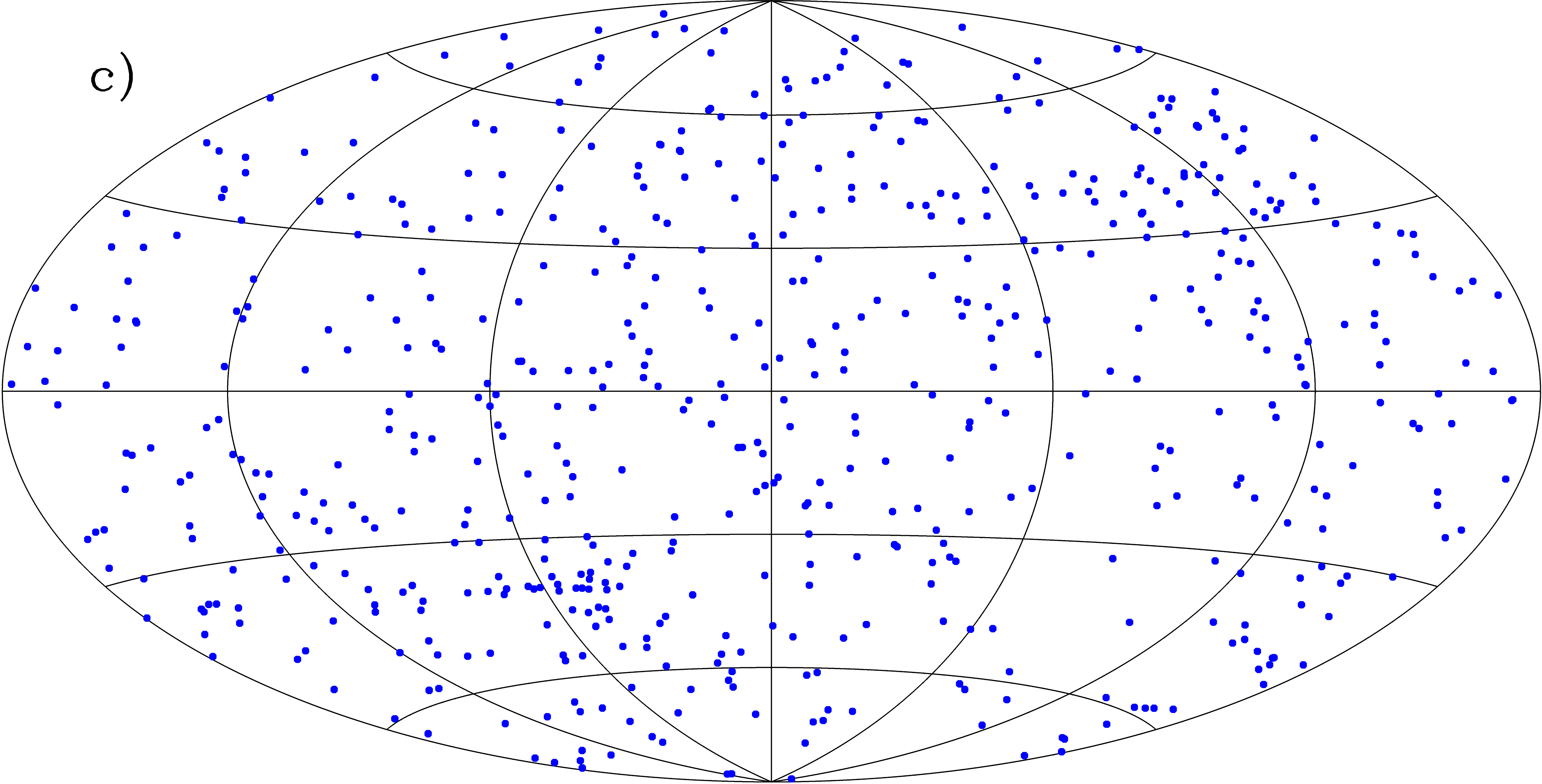}
   \includegraphics[width=73mm]{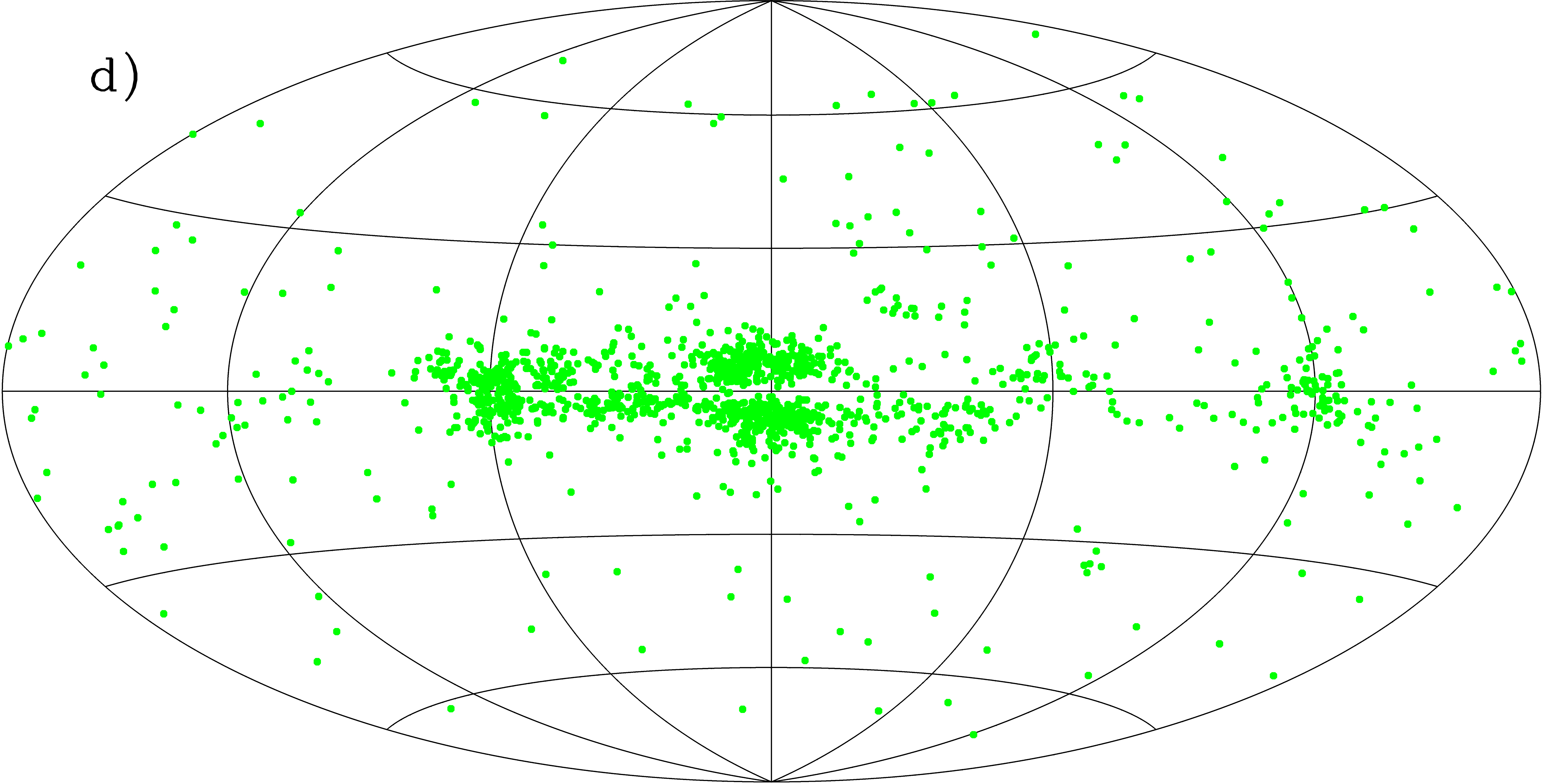}
   \includegraphics[width=73mm]{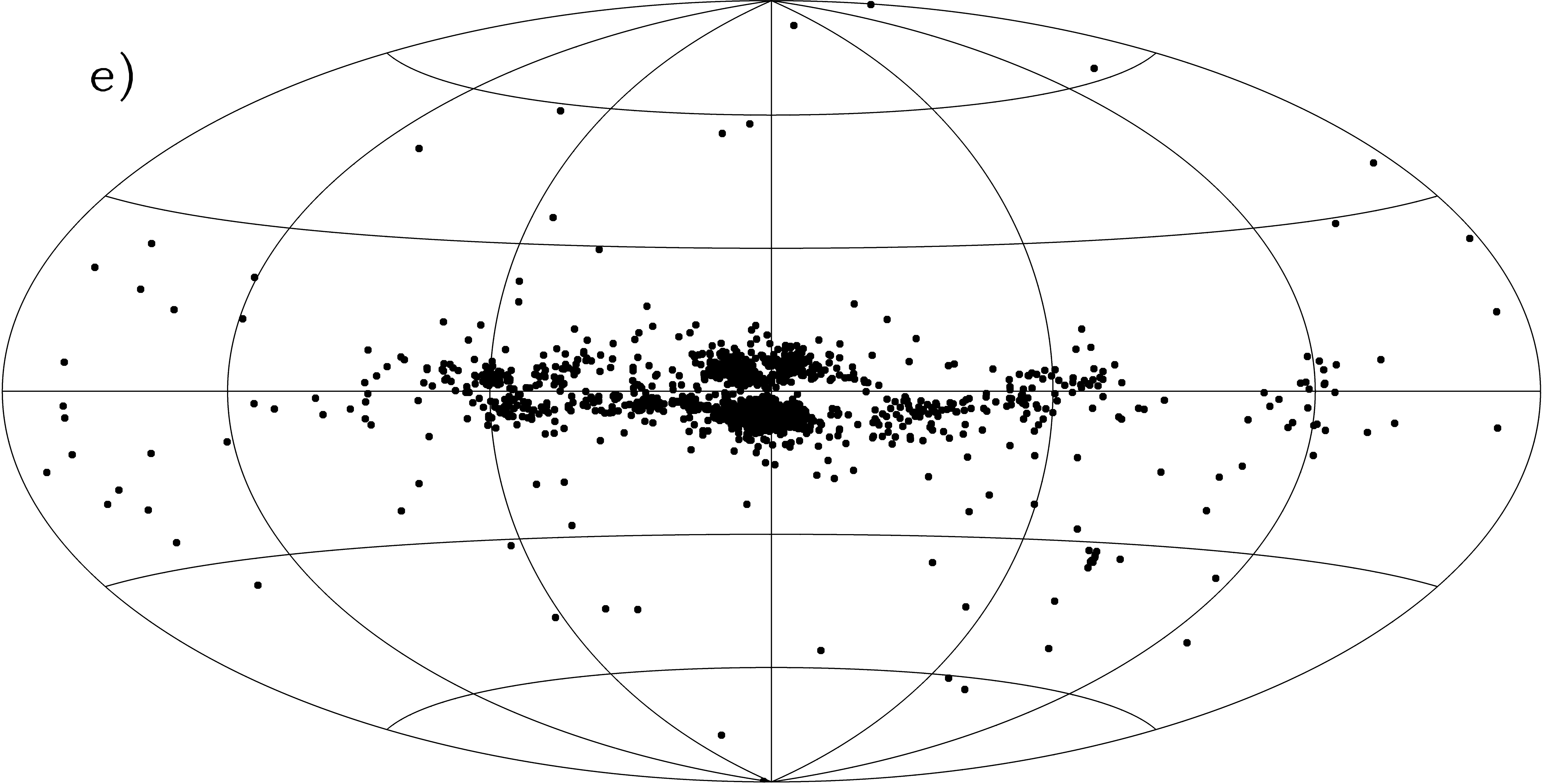}
      \caption{Sky distributions in Galactic coordinates $l,b$ 
              (centred on $l=0\degr, b=0\degr$, with $l$ rising 
              to the left, Galactic north pole $b=90\degr$ is up) 
              of all objects
              in the {\it Gaia} DR2 20\,pc sample.
              Panels a) (with red symbols),
              b) (magenta), c) (blue), d) (green), and e) (black) 
              correspond to
              subsamples A-E, respectively.
              }
      \label{Fig_skyq}
   \end{figure}

As we are in particular interested in the location of faint objects and 
possible nearby UCDs falling in crowded fields in {\it Gaia} DR2 CMDs, 
we also estimated their photometric quality. 
For ''good'' photometry, all of the following three criteria had to be
fulfilled (otherwise the photometry was considered ''bad''):
\begin{description}
\item[phot\_q1] - \textsf{phot\_bp\_mean\_flux\_over\_error}$>10$,
\item[phot\_q2] - \textsf{phot\_rp\_mean\_flux\_over\_error}$>10$,
\item[phot\_q3] - \textsf{phot\_bp\_rp\_excess\_factor} $<1.3+0.06\times(BP-RP)^2$,
\end{description}
where the first two conditions ensure better than 10\% uncertainty 
in $BP$ and $RP$ fluxes corresponding to $\approx$0.1\,mag in
$BP$ and $RP$ magnitudes \citepads{2018A&A...616A...2L} and
were also applied in ''Selection A'' of \citetads{2018A&A...616A...2L}.
The third condition, applied in ''Selection C'' 
of \citetads{2018A&A...616A...2L} and 
by \citetads{2018A&A...616A..10G},
uses the flux excess factor indicating problematic 
$BP$ or $RP$ photometry 
\citepads{2018A&A...616A...4E}, mainly of faint sources in
crowded fields.

   \begin{figure}
   \centering
   \includegraphics[width=\hsize]{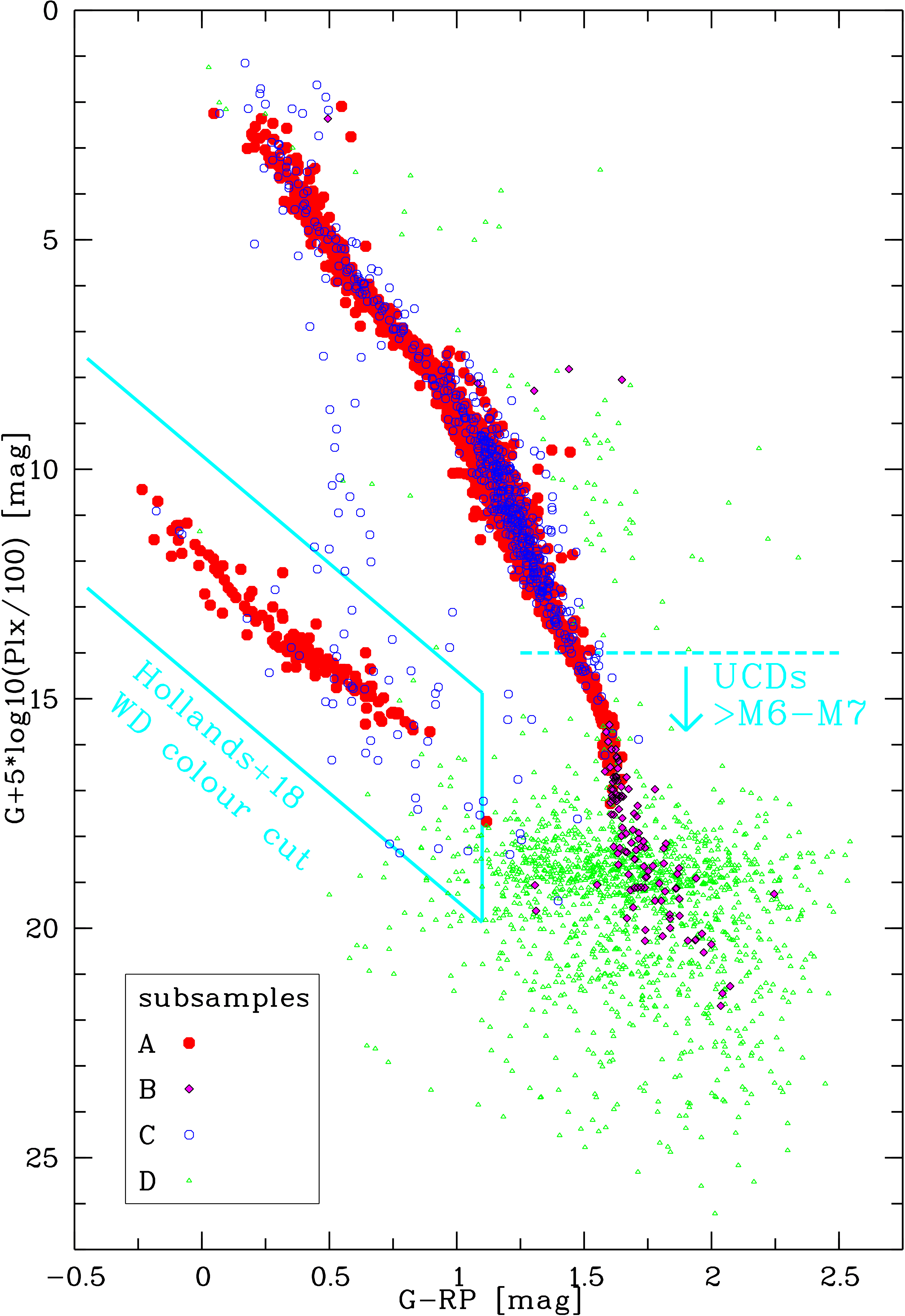}
      \caption{CMD of the full {\it Gaia} DR2 20\,pc sample.
              The subsamples A-D defined in Sect.\ref{SubSect_q}
              are shown with the same colours (red, magenta, blue,
              and green) as in Fig.\ref{Fig_Ghist}. The
              objects with good astrometry (subsamples A and B) 
              are plotted as filled (large and small)
              symbols, those with bad astrometry (subsamples C and D) 
              as open (large and small) symbols.
              Here, and in Fig.\ref{Fig_apm}, 
              the magenta filled symbols are overplotted with 
              black open lozenges for better visibility.
              The colour cuts applied for WDs by
              \citetads{2018MNRAS.480.3942H} and the absolute
              magnitude limit of $M_G>14$\,mag used in this study for
              selecting UCDs of spectral types $>$M6-M7 are drawn
              in the background as cyan thick solid and dashed lines, 
              respectively.
              }
      \label{Fig_CMDall}
   \end{figure}

Out of 4065 objects with available $BP$$-$$RP$ colours in the {\it Gaia} DR2
20\,pc sample, only 48\% are qualified by us to have good astrometry. Their
$G$ magnitude distribution (subsamples A and B) is shown in 
Fig.~\ref{Fig_Ghist}. Only 3\%, mainly at faint magnitudes, have good 
astrometry but bad photometry (subsample B). The histogram of subsample C, 
with bad astrometry but good photometry, peaks at a similarly bright 
magnitude of $G$$\approx$12\,mag as that of subsample A. Subsample D, with 
both bad astrometry and bad photometry, contains more than one third (38\%)
of all objects and shows a strong concentration at faint 
$G$ magnitudes, similar to the additional 2040 objects without
$BP$$-$$RP$ colour measurements (subsample E), also shown
in Fig.~\ref{Fig_Ghist}. Only a negligible part of subsample E
(10 objects) can be qualified as objects with good astrometry (not shown
in Fig.~\ref{Fig_Ghist}).

The subsamples of different astrometric and photometric quality show
characteristic distributions on the sky (Fig.\ref{Fig_skyq}). 
The two top panels a) and
b) display the objects with good astrometry. They are, as
expected for nearby objects, uniformely 
distributed over the sky, except for some empty regions seen in panel a),
where only few observations could be included in {\it Gaia} DR2 
(\textsf{visibility\_periods\_used}$<9$).
Panel c) corresponding to bad astrometry and good 
photometry logically contains overdensities in those regions
that were 
empty in panel a). Other 
concentrations of objects connected to the scanning
history of {\it Gaia} DR2 are not seen but become visible
when we plot higher numbers of these objects within a larger distance 
limit of 100\,pc (not shown here).
Panel d) with both
bad astrometry and bad photometry displays a very high
density in the crowded regions expected close to the 
Galactic centre and along 
the Galactic plane, and in the Large Magellanic cloud
(at $l\approx280\degr, b\approx-33\degr$).
Finally, panel e) represents objects 
without $BP$$-$$RP$ colours
with a sky distribution very similar to that of panel d).
   
The CMD in Fig.\ref{Fig_CMDall} represents the full {\it Gaia} DR2 20\,pc 
sample of 4108 objects, except for one bright star (\object{Hip 65109})
with bad photometry and bad astrometry 
falling outside the frame.  
Subsample A, with good astrometry and good photometry, shows a
relatively clean MS as well as a clear WD sequence located in
the central part of the WD colour box defined by
\citetads{2018MNRAS.480.3942H}. In particular, the lower MS,
in the range 13\,mag$<M_G<$17\,mag is relatively narrow, if we
consider the red large filled symbols only. At fainter absolute magnitudes
most objects with good astrometry have bad photometry and belong
to subsample B (magenta small filled symbols). Only a few brighter 
objects of subsample B 
appear to the right of the MS. 
The majority of objects with 
bad astrometry but good photometry (subsample C) follow the MS, 
but there are also many of the corresponding blue large open symbols 
scattered below the upper MS towards and around the WD region. 
Finally, subsample D,
representing bad astrometry and bad photometry, mainly separates into 
two wide distributions, one very wide (the majority of
green small open symbols) 
scattered around the bottom of the MS and reaching the WD colour box
of \citetads{2018MNRAS.480.3942H}, 
and one (with fewer objects) 
scattered to the right side of the MS above the UCD 
absolute magnitude cut, which we define at $M_G=14$\,mag.

The faintest objects with good astrometry in our {\it Gaia} DR2 20\,pc 
sample (magenta small filled symbols in Fig.\ref{Fig_CMDall})
have absolute magnitudes $M_G<22$\,mag, comparable with the value
previously derived for mid-T dwarfs 
by \citetads[][see their Fig.~15]{2017MNRAS.469..401S}
based on {\it Gaia} DR1 \citepads{2016A&A...595A...2G} photometry 
and parallaxes from the literature.
On the other hand, our more problematic objects with bad astrometry and 
bad photometry (green small open symbols) reach a faint limit 
of $M_G\approx$26\,mag that would probably correspond to late-T or even 
early-Y dwarfs, if real. In the following, we will show, which
of the many faint data points at the bottom of the MS in the {\it Gaia} 
DR2 20\,pc CMD (Fig.\ref{Fig_CMDall}) correspond to the relatively small
numbers of already known and new real UCDs, respectively. 

\subsection{known UCDs and WDs}
\label{SubSect_ucdwd}

After their initial study of known 
spectroscopically classified UCDs in {\it Gaia} 
DR1 \citepads{2017MNRAS.469..401S}, \citetads{2019MNRAS.485.4423S}
investigated the structure at the bottom of the MS based on
{\it Gaia} DR2. The input catalogues of these two studies
included 1885 and 3093 known M, L, T, and Y dwarfs, respectively. 
Since the larger list is dominated by L and T dwarfs, which are
mostly too faint for {\it Gaia}, only 695 of the 3093 UCDs were found 
by \citetads{2019MNRAS.485.4423S} in 
{\it Gaia} DR2, and only 134 fall in our {\it Gaia} DR2 20\,pc sample. 
Concerning potential members of the {\it Gaia} DR2 20\,pc sample, only 
few additional objects were added to the input catalogue 
of \citetads{2019MNRAS.485.4423S}. Among them, there were three mid-
to late-L dwarf candidates with estimated photometric distances 
between 10\,pc and 20\,pc discovered as red 
high proper motion (HPM)
objects from the 
Two Micron All Sky Survey \citepads[2MASS;][]{2006AJ....131.1163S}, the 
Wide-field Infrared Survey Explorer \citepads[WISE;][]{2010AJ....140.1868W},
and with help from {\it Gaia} DR1 and other catalogues
\citepads{2018RNAAS...2...33S}. All three were 
subsequently spectroscopically confirmed 
\citepads{2018ApJ...868...44F,2018RNAAS...2...50C,2019RNAAS...3...30L},
but only the nearest one (\object{2MASS J19251275+0700362}) has 
a measured {\it Gaia} DR2 parallax.

   \begin{figure*}
   \centering
   \includegraphics[width=18cm]{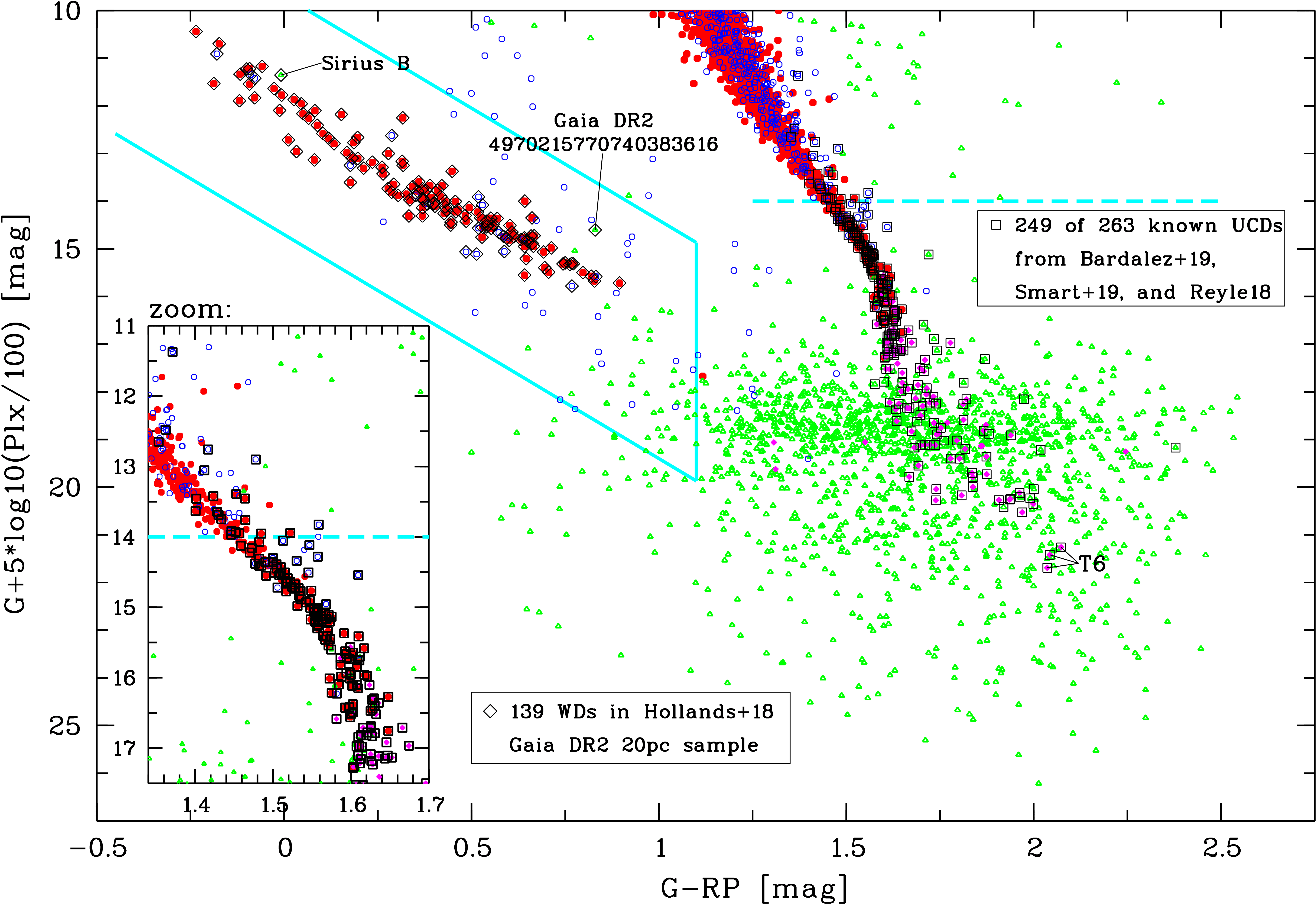}
      \caption{Faint part of the CMD, with a zoom into the
              MS region around our UCD limit (dashed line), 
              of the full {\it Gaia} DR2 20\,pc sample
              with known WDs from \citetads{2018MNRAS.480.3942H}
              overplotted as open lozenges
              and known UCDs from
              \citetads{2019ApJ...883..205B,2019MNRAS.485.4423S},
              and \citetads{2018A&A...619L...8R} overplotted
              as open squares. Other 
              coloured symbols 
              and lines are the same
              as in Fig.~\ref{Fig_CMDall}
              Two WDs with problematic astrometry and photometry
              and the three latest-type (T6) known UCDs are marked.
              }
      \label{Fig_CMDallknown}
   \end{figure*}

\citetads{2019MNRAS.485.4423S} also mentioned the 
VVV Infrared Astrometric Catalogue \citepads[VIRAC;][]{2018MNRAS.474.1826S} 
as a source for new nearby UCD candidates. The latter work includes
ten objects with VIRAC parallaxes $>$40\,mas. We found all of them in
{\it Gaia} DR2, but only four falling in
the {\it Gaia} DR2 20\,pc sample, 
including one WD (\object{VVV J14115932-59204570})
and one object that appears slightly brighter
than our chosen UCD absolute magnitude cut of $M_G>14$\,mag.
The remaining two UCD candidate members of the {\it Gaia} DR2 20\,pc 
sample from \citetads{2018MNRAS.474.1826S} are 
the L5 dwarf \object{VVV J17264015-27380372} formerly discovered
as \object{VVV BD001} by \citetads{2013A&A...557L...8B} and
included in \citetads{2019MNRAS.485.4423S},
and another mid-L dwarf candidate (with similar $M_G$$\approx$16.7\,mag
and colour $G$$-$$RP$$\approx$1.7\,mag), \object{VVV J17134060-39521521},
that is missing in \citetads{2019MNRAS.485.4423S}.

\citetads{2018A&A...619L...8R} searched for new UCD candidates in
{\it Gaia} DR2 and found 14\,915 objects, all with 
available $G$$-$$RP$ colours, and provided spectral type estimates
based on photometry. According to these estimates, given in the
corresponding machine-readable table including only few of
the original {\it Gaia} DR2 data (without errors), most of the 
candidates are late M dwarfs, whereas only 280 ($<$2\%) 
are L dwarfs. After an initial selection of only 74 out of
14\,915 objects with parallaxes $>$40\,mas, we used their parallaxes 
and $G$ magnitudes for a cross-match with the full {\it Gaia} DR2,
since the {\it Gaia} DR2 names were not correct for many
sources in the machine-readable table of \citetads{2018A&A...619L...8R}.
Finally, we selected
25 of the 74 objects falling in our {\it Gaia} DR2 20\,pc sample,
according to our soft parallax cut involving
the parallax errors (see beginning of Sect.~\ref{Sect_20pc}).
There was only one object in common with the above mentioned subsample
of 134 objects from \citetads{2019MNRAS.485.4423S}, and the
missing \object{VVV J17134060-39521521} was found among those
25 objects from \citetads{2018A&A...619L...8R}. These 25 UCD
candidates are systematically brighter in absolute magnitudes
(13.7\,mag$<M_G<$17.4\,mag, with a peak at $M_G$$\approx$15\,mag)
than the 134 objects from \citetads{2019MNRAS.485.4423S}
(14.5\,mag$<M_G<$21.7\,mag, with a peak at $M_G$$\approx$17\,mag).

   \begin{figure}
   \centering
   \includegraphics[width=\hsize]{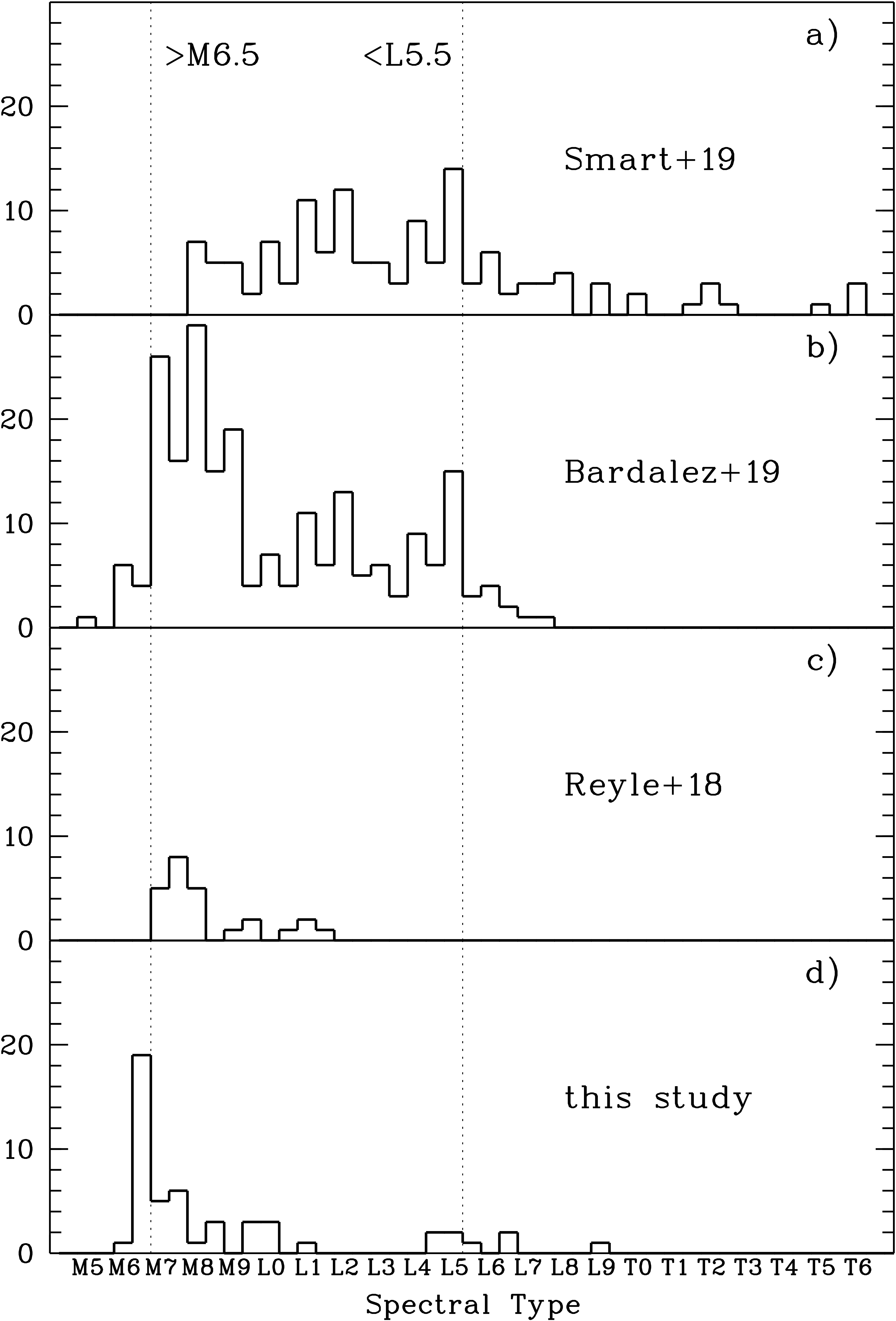}
      \caption{Distribution of spectral types in the
              {\it Gaia} DR2 20\,pc subsamples of known
              spectroscopically classified UCDs 
              from
              \citetads{2019MNRAS.485.4423S} (panel a)) and
              \citetads{2019ApJ...883..205B} (panel b)).
              Panels c) and d) show the distributions for
              the UCD candidates with photometrically 
              estimated spectral types in \citetads{2018A&A...619L...8R}
              and this study, respectively. The dotted lines mark the
              boundaries of the volume-limited spectroscopic sample
              of \citetads{2019ApJ...883..205B} between M6.5 and M7 
              (also used as the
              classical UCD dividing line) and L5 and L5.5 dwarfs.
              }
      \label{Fig_SpThis}
   \end{figure}

A volume-limited spectroscopic sample of M7-L5 dwarfs within 25\,pc
was presented by \citetads{2019ApJ...883..205B}. Out of the
471 entries in their (machine-readable) Table~6, we found 220
belonging to the {\it Gaia} DR2 20\,pc sample, of which however
only 203 were listed with their {\it Gaia} DR2 parallaxes 
by \citetads{2019ApJ...883..205B}. The 220 entries correspond to
216 {\it Gaia} DR2 sources (four are unresolved in {\it Gaia} DR2).
From these 216 UCDs, there are 109 in common with
\citetads{2019MNRAS.485.4423S}, whereas \citetads{2019ApJ...883..205B}
did not include 24 nearby UCDs of \citetads{2019MNRAS.485.4423S}, 
mainly because 
these have later spectral types (L7-T6). Only two objects
out of the 25 new UCD candidates from \citetads{2018A&A...619L...8R} 
with photometrically estimated spectral types were
found among the 216 UCDs of \citetads{2019ApJ...883..205B} 
(one of them was also found in \citetads{2019MNRAS.485.4423S}).
The 216 {\it Gaia} DR2 sources of \citetads{2019ApJ...883..205B}
show a wide distribution in absolute magnitudes
(11.3\,mag$<M_G<$20.3\,mag, with a peak at $M_G$$\approx$15\,mag).
The peak in their absolute magnitude distribution is similar to
that of the much smaller sample of \citetads{2018A&A...619L...8R},
but the even brighter border of the magnitude interval is more typical of 
mid-M dwarfs than of UCDs. However, there is a large spread in absolute
magnitudes of M7 dwarfs. Using the spectral types given by
\citetads{2019ApJ...883..205B} 
and selecting only well-measured (resolved 
or single) objects in {\it Gaia} DR2, we find mean $M_G$ values
(and standard deviations) of 14.04 (0.470)\,mag for six M6.0 
and 14.02 (1.01)\,mag for 26 M7.0 dwarfs. For later types we see
a clear trend towards fainter absolute magnitudes: 15.12 (0.48)\,mag
for 26 M8.0 and 16.07 (0.27)\,mag for 19 M9.0 dwarfs. These
numbers confirm that our chosen UCD absolute magnitude limit of
$M_G>14$\,mag, 
indicated by a dashed line in all our CMDs,
roughly corresponds to M6-M7 spectral types.

The {\it Gaia} DR2 20\,pc sample of \citetads{2019ApJ...883..205B} 
includes
some M5.0-M6.5 dwarfs, which do not meet the classical UCD definition by
spectral types $\gtrsim$M7 (Fig.\ref{Fig_SpThis}). Within the spectral 
type range
M7-L5 of their volume-limited sample (marked by dotted lines), their
L dwarfs mostly overlap with those in \citetads{2019MNRAS.485.4423S}, 
whereas they added many more M-type UCDs. 
The UCD candidates of \citetads{2018A&A...619L...8R} were selected
by photometrically estimated spectral types $>$M6.5 corresponding
to the spectral type definition of UCDs. However, their spectroscopic
classification is still needed and some of these objects may in fact
be of earlier spectral types. On the other hand, \citetads{2018A&A...619L...8R}
may have excluded some real UCDs by their photometric spectral type cut.
The spectroscopic classification
of the new UCD candidates of our study will be of course necessary, too.
This may lead to changes both towards earlier and later spectral types.

In total, we identified 263 known UCDs from \citetads{2019ApJ...883..205B},
\citetads{2019MNRAS.485.4423S}, and \citetads{2018A&A...619L...8R}
in the {\it Gaia} DR2 20\,pc sample. Out of these 263 objects, 14
lack the $BP$$-$$RP$ and $G$$-$$RP$ colour measurements and belong 
to subsample E. The other 249 known
UCDs are overplotted in Fig.~\ref{Fig_CMDallknown} as open squares.
They are wide-spread with respect to their astrometric and
photometric quality criteria defined in Sect.~\ref{SubSect_q}.
Out of these 249 known UCDs, 108 (43\%) fall in subsample A,
94 (38\%) in subsample B, 22 (9\%) in subsample C, and 25 (10\%) in
subsample D.

Concerning the nearby WD census, \citetads{2018MNRAS.480.3942H}
have already investigated the {\it Gaia} DR2 20\,pc sample 
and found nine additions, the closest of which lies at a
distance of 13\,pc, based on {\it Gaia} DR2 parallaxes.
Their updated WD sample consists of 139 objects. According to
their study of the 
distance-dependent 
{\it Gaia} DR2 completeness,
some very nearby WDs (e.g. four known WDs within 10\,pc)
are still missing, whereas {\it Gaia} DR2 should be close to
complete with respect to WDs at about 20\,pc.
In all our CMDs we have marked 
the WD colour cuts used by \citetads{2018MNRAS.480.3942H}. 
Out of the 139 known WDs from \citetads{2018MNRAS.480.3942H},
overplotted in Fig.~\ref{Fig_CMDallknown} as open lozenges,
119 (86\%) belong to subsample A, none to subsample B, 18 (13\%)
to subsample C, and only 2 (1\%) to subsample D. These numbers
and the fact that all 139 WDs have measured $G$$-$$RP$ colours 
confirm the higher quality of the {\it Gaia} DR2 data of nearby WDs
compared to that of nearby UCDs. This is simply related to
the relatively bright $G$ magnitudes and blue $G$$-$$RP$ colours,
at which nearby WDs can be observed. 

However, nearby
WD companions in close binary systems and nearby WDs overlapping
with background objects (e.g. in crowded fields along the Galactic plane)  
may be difficult targets for {\it Gaia}. The two known WDs with
both bad astrometry and bad photometry (subsample D) are labelled
in Fig.\ref{Fig_CMDallknown}. Whereas one of them, \object{Sirius B},
was obviously hard to be measured by {\it Gaia}, the other,
\object{Gaia DR2 4970215770740383616}, is one of the nine new
WDs found by \citetads{2018MNRAS.480.3942H}. This is a previously
known 
HPM star 
(= \object{LP 941-19}) at high
Galactic latitude ($b\approx-71\degr$) that happened
to move at the {\it Gaia} DR2 epochs 
very close to a similarly bright 
and blue background star (angular
separation $\approx$1.2\,arcsec).

\subsection{search for new UCDs and WDs}
\label{SubSect_search}

\begin{figure*}
   \centering
   \includegraphics[width=18cm]{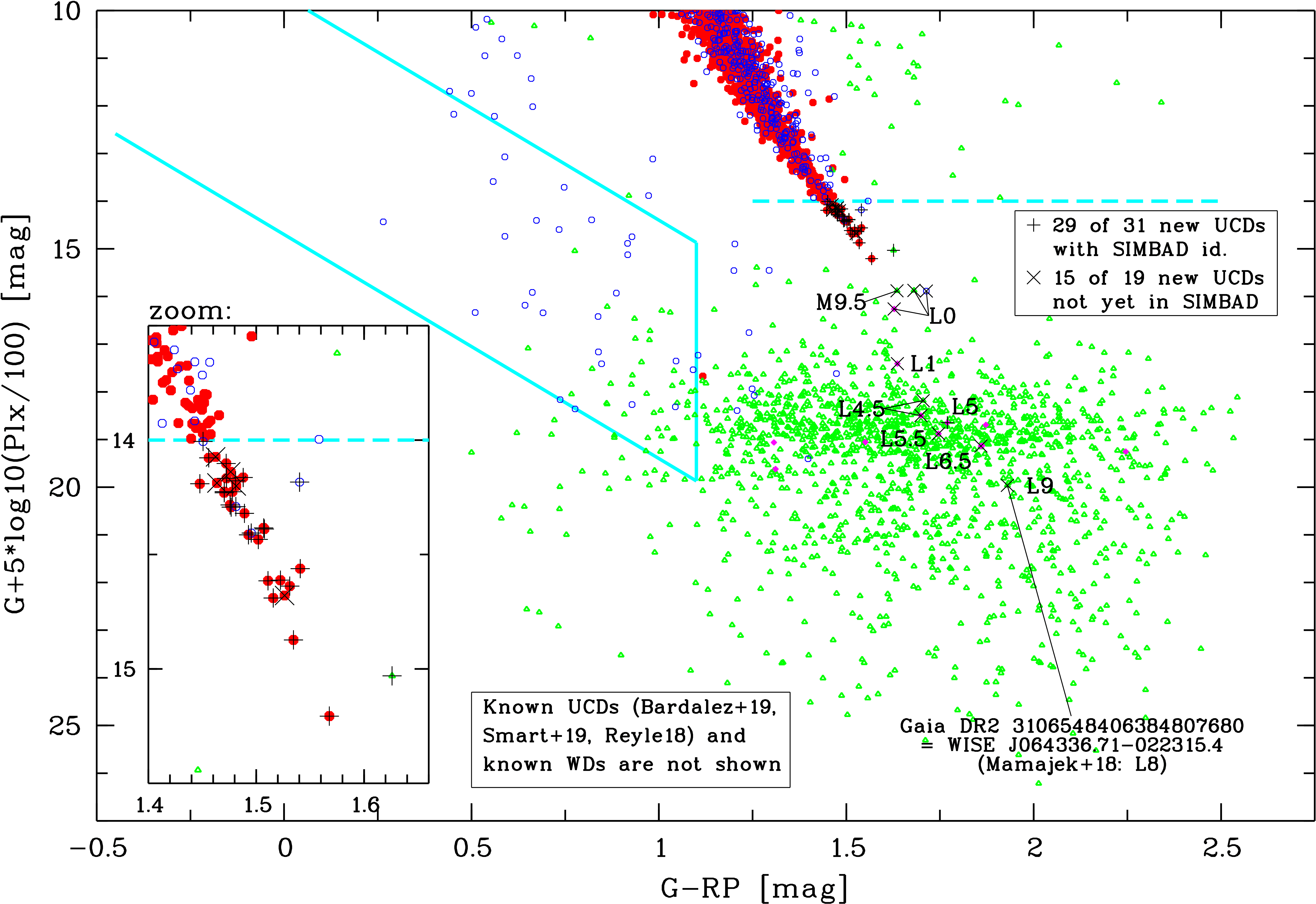}
      \caption{Faint part of the
              CMD of the {\it Gaia} DR2 20\,pc sample after
              excluding 139
              known WDs from \citetads{2018MNRAS.480.3942H}
              and 263 known UCDs from
              \citetads{2019ApJ...883..205B,2019MNRAS.485.4423S},
              and \citetads{2018A&A...619L...8R}.
              A closer zoom in the
              MS region at and below our UCD magnitude limit 
              (dashed line) is also shown. Pluses (for
              objects with SIMBAD identification) and crosses
              (for objects not yet found in SIMBAD)
              mark the 44 of our 50 new UCDs
              that have $G$$-$$RP$ colours.
              Other 
              coloured symbols and lines are as 
              in Fig.~\ref{Fig_CMDallknown}.
              The latest-type new UCDs are marked by their
              photometrically determined spectral 
              types (Sect.\ref{Sect_photclass}).
              }
      \label{Fig_CMDallnoknown}
   \end{figure*}

After excluding all known UCDs and WDs described in
Sect.~\ref{SubSect_ucdwd} from the 
{\it Gaia} DR2 20\,pc sample, we tried to find 
additional UCDs 
(called ''new UCDs'' in the following)
and cool WDs among all $\approx$3500
remaining objects with absolute magnitudes $M_G>14$\,mag.
Only 44\% of these candidates have $G$$-$$RP$ colour
measurements, and their CMD is presented 
in Fig.~\ref{Fig_CMDallnoknown}). At $M_G>14$\,mag,
the CMD is clearly dominated by candidates with bad astrometry 
and bad photometry,
which are plotted as green small open triangles. 
There is only a small number of
candidates with good astrometry, 
which are shown as red and magenta filled symbols.
The corresponding UCD candidates are 
concentrated at the faint end of the MS in the
interval 14\,mag$<M_G<$15\,mag and 
they also extend 
to fainter magnitudes with a larger scatter in
colour. 
This distribution is similar to that of the
previously known UCDs shown in Fig.~\ref{Fig_CMDallknown}. 
Only one WD candidate with good astrometry and
good photometry falls close to the border of the WD colour box
of \citetads{2018MNRAS.480.3942H} at $G$$-$$RP$$\approx$1.1\,mag.
Within this colour box, there is no more trace of a WD sequence. In
addition to our 
general selection of UCD and WD candidates
with $M_G>14$\,mag, we also searched for hotter WDs among the small
number of brighter objects
with $G$$-$$RP$$<$1\,mag 
falling within or a few magnitudes above the WD colour box 
of \citetads{2018MNRAS.480.3942H}. All these objects have bad astrometry
and are shown with open symbols.

The {\it Gaia} DR2 5-parameter astrometric solution includes both
the parallax and the proper motion of a star. The latter can
be easily compared with independent measurements and provide a
hint on the reliability of the former. Therefore,
we checked the given {\it Gaia} DR2 proper motions
of all candidates by 
\begin{description}
\item[step 1] - visual inspection of multi-epoch finder charts 
from the NASA/IPAC Infrared Science Archive (IRSA;
http://irsa.ipac.caltech.edu/applications/finderchart/),
\item[step 2] - comparison with other accurate proper motion catalogues
provided by VizieR (http://vizier.u-strasbg.fr/) if available, and
\item[step 3] - 
comparison with our own proper motion determination 
using positional data from various catalogues, if needed.
\end{description}

In step 1 we took advantage of the near-infrared (NIR) images from
2MASS (epoch $\approx$2000) and WISE (epoch $\approx$2010), which
are provided by IRSA at a glance with the multi-colour
optical images (epochs between 1950 and 2000) from photographic
Schmidt plates and, in part of the sky, from the Sloan Digital Sky
Survey \citepads[SDSS;][]{2009ApJS..182..543A}.
For UCDs we expected an increase in brightness from the optical
to the NIR so that all nearby UCDs should be well seen
in 2MASS and WISE images. On the other hand, nearby WDs may be
visible or not in WISE and 2MASS but should appear bright and
relatively blue in the optical.

In step 2 we found the catalogue of \citetads{2017ApJS..232....4T}
very useful. This proper motion catalogue combined Gaia DR1,
the Pan-STARRS1 Surveys \citepads[PS1;][]{2016arXiv161205560C},
SDSS, and 2MASS astrometry over about 3/4 of the sky. For some
of our brighter late-M candidates other proper motion catalogues
like the PPMXL \citepads{2010AJ....139.2440R} and the first
U.S. Naval Observatory Robotic Astrometric Telescope Catalog
\citepads[URAT1;][]{2015AJ....150..101Z} could also be used for
comparison with the {\it Gaia} DR2 proper motions.

We performed step 3, when it was necessary, 
in particular in the
case of red objects (which are brighter in 2MASS and WISE)
with small proper motions 
according to both {\it Gaia} DR2
and our visual check in step 1. For our own proper motion
determination we combined {\it Gaia} DR1 and DR2 positions
with those from PS1, 2MASS and WISE. When available, we
also included positional measurements from
the UKIDSS Galactic Plane Survey \citepads[GPS;][]{2008MNRAS.391..136L},
the VISTA Hemisphere Survey \citepads[VHS;][]{2013Msngr.154...35M},
the fifth U.S. Naval Observatory Astrograph
Catalogue \citepads[UCAC5;][]{2017AJ....153..166Z},
the final Carlsberg Meridian
Catalogue \citepads[CMC15;][]{2014AN....335..367M}, and
the first SkyMapper Southern Survey \citepads{2018PASA...35...10W}.

   \begin{figure*}
   \centering
   \includegraphics[width=17cm]{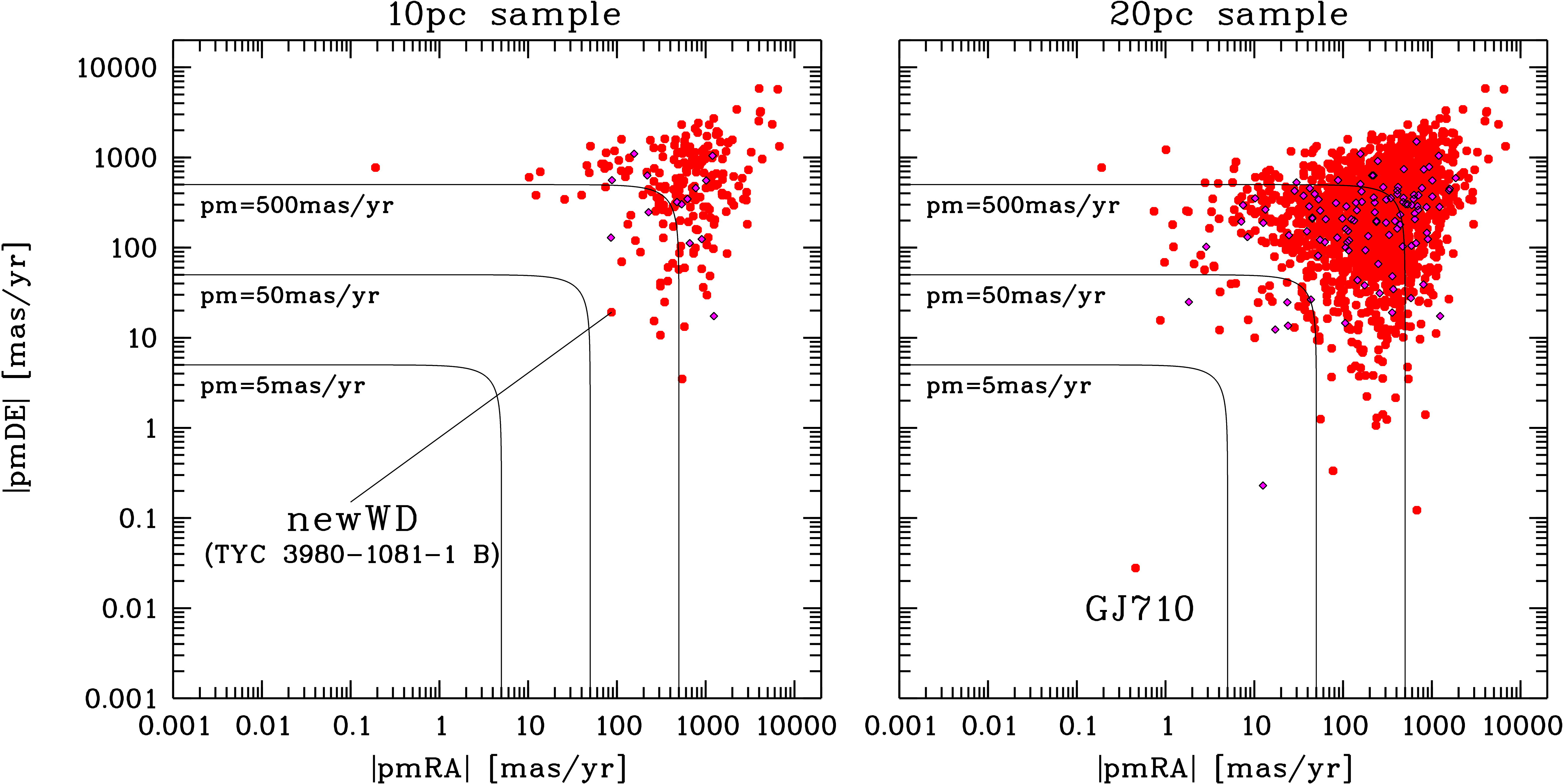}
   \includegraphics[width=17cm]{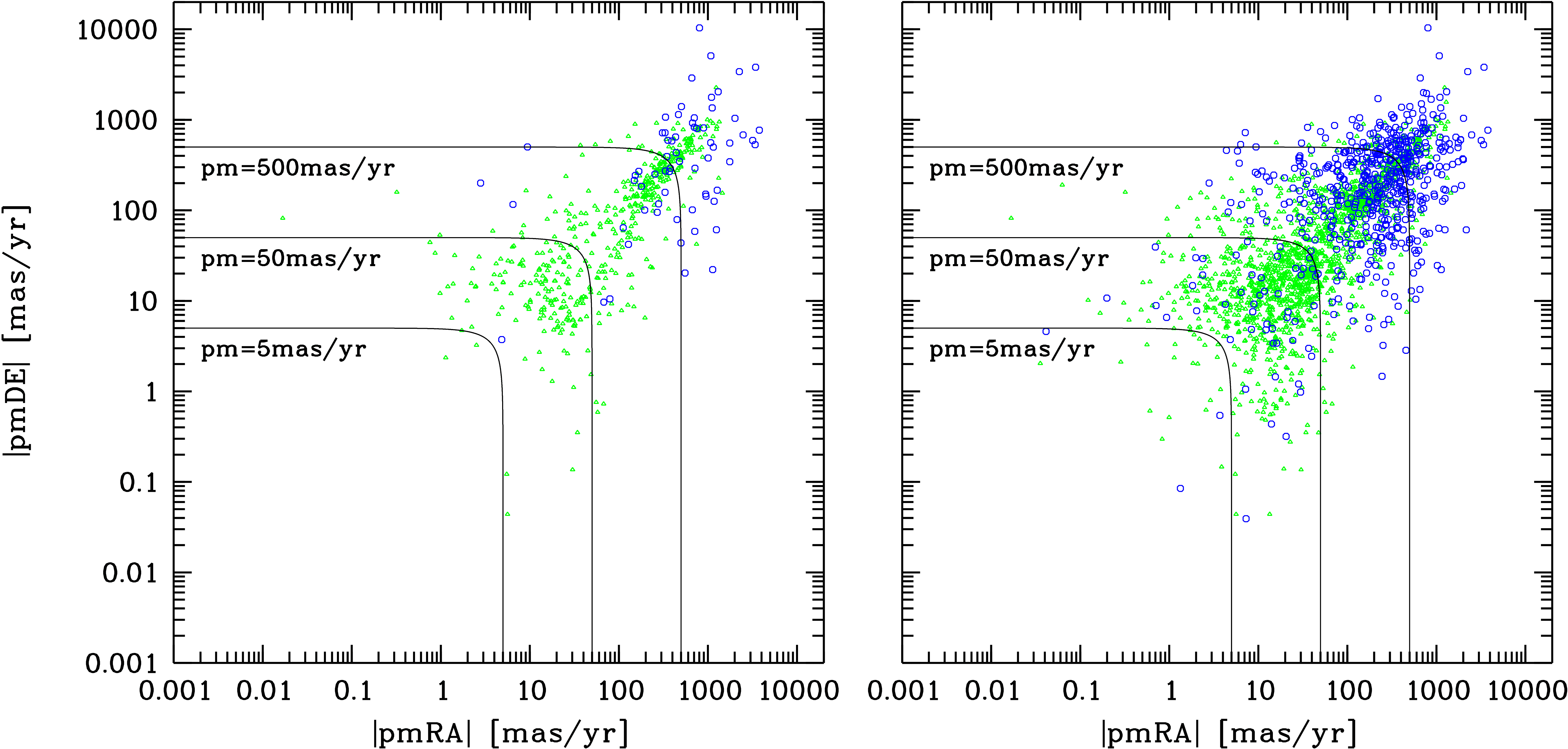}
      \caption{Absolute proper motion components $|pmRA|$
              and $|pmDE|$ of stars in the {\it Gaia} DR2 10\,pc
              (left panels) and 20\,pc samples (right panels)
              are plotted in logarithmic scale.
              Total proper motions of 5\,mas/yr, 50\,mas/yr,
              and 500\,mas/yr are indicated by solid lines.
              The symbols for the subsamples A and B (top row)
              and C and D (bottom row) are the same as in
              Fig.\ref{Fig_CMDall}. 
              Two well-measured nearby stars with the
              smallest proper motions are marked in the top
              row.}
      \label{Fig_apm}
   \end{figure*}

The smaller the {\it Gaia} DR2 proper motion, the more difficult 
it is to confirm or reject it from comparison with other data.
However, among the known objects within 20\,pc there are only few
with very small proper motions. In Fig.\ref{Fig_apm}
the absolute proper motion components $|pmRA|$ and $|pmDE|$
are plotted for our subsamples A-D of different astrometric
and photometric quality. If we consider only the objects with
good astrometry in the top row, the majority of the nearest 
objects within 10\,pc (top left) 
are clear HPM objects that
have total proper motions $>$500\,mas/yr.
Their smallest proper motion is $\lesssim$100\,mas/yr, and we note
that the corresponding object, 
the new WD \object{TYC 3980-1081-1 B}, was only recently discovered
\citepads{2018A&A...613A..26S}. Extending the distance of the
objects with good astrometry to 20\,pc (top right), we see the
majority having total proper motions $>$100\,mas/yr, while
few objects have smaller proper motions but still mostly $>$10\,mas/yr.
Interestingly, there is only one exception of a proper motion 
of $<$1\,mas/yr, which  belongs to the well-known 
K dwarf \object{GJ 710} that will have a close encounter with the 
Sun 
in about 1.3\,Myr \citepads{2015A&A...575A..35B}.

When we look at the bottom row in Fig.\ref{Fig_apm},
containing the objects with bad astrometry, we see two
remarkable differences that mainly concern the objects with
both bad astrometry and bad photometry (subsample D;
green small open symbols).
First, many of these objects have smaller proper motions than
the astrometrically good objects shown in the upper row. Second,
there is a suspicious trend to an equal (large) size of both proper 
motion components (a concentration of objects along the upper 
diagonal of the plots), which would not be expected for real proper 
motion distributions. We speculate that this could be the result of 
systematic mismatches of objects in crowded fields that
happened during the data reduction of {\it Gaia} DR2.
Such errorneous 
HPM 
components that are equal 
in size can also be found for distant objects with very small
parallaxes. For instance, we found 50\% of the sample of 28 distant
hyper-velocity candidates of \citetads{2019ApJS..244....4D}, which 
appear as {\it Gaia} DR2
HPM stars 
in the Galactic
centre and anticentre direction, to be affected by this error.

Although we observed these systematic differences in the
proper motions of astrometrically good and bad objects, we 
tried to find real UCDs and WDs among \textit{all} our candidates
using our three-step procedure for checking the {\it Gaia} DR2
proper motions. The success rate of finding new UCDs and WDs
was expected to be very low in the Galactic plane regions,
dominated by objects of subsamples D and E. Because the DR2 20\,pc
WD sample was already filled by \citetads{2018MNRAS.480.3942H}
and contained predominantly well-measured objects of
subsample A, we did not expect to find many new nearby WDs.
On the other hand, the fact that a relatively large fraction 
of the known UCDs are members of subsample D or belong to
subsample E as they have no
measured $G$$-$$RP$ colours (Sect.~\ref{SubSect_ucdwd})
supported our motivation to check the complete sample of
unknown UCD and WD candidates.

We were able to confirm 50 of our candidates as red 
HPM objects, 
mostly with well-measured bright 
counterparts in 2MASS and also appearing bright
but sometimes merged with other objects in WISE. We
considered all these objects as new UCD members of 
the {\it Gaia} DR2 20\,pc sample (Fig.~\ref{Fig_CMDallnoknown};
six out of 50 are not shown 
because of they lack $G$$-$$RP$ colours). 
On the other hand, we did not succeed in finding new 
WD members. In particular, the 
only one object falling at the right border of the WD colour box
and belonging to subsample A in Fig.~\ref{Fig_CMDallnoknown}
turned out to have a doubtful proper motion in {\it Gaia} DR2.

Our new UCDs were not included in the three
previous studies of UCDs based on {\it Gaia} DR2 data
\citepads{2019ApJ...883..205B,2019MNRAS.485.4423S,2018A&A...619L...8R}.
However, these studies may have missed some nearby UCDs with
previous non-{\it Gaia} distance estimates. Therefore,
after looking in the SIMBAD data base, we divided our 
50 new UCDs in two groups, one with SIMBAD identifications
(Table~\ref{Tab_newUCDinS}, pluses in Fig.~\ref{Fig_CMDallnoknown}) 
and one without (Table~\ref{Tab_newUCDnoS}, crosses in
Fig.~\ref{Fig_CMDallnoknown}). 

The first group consists of 31 objects
previously found as nearby candidates in
various 
HPM surveys, 
as seen from their designations (e.g. 
LSPM \citepads{2005AJ....129.1483L},
2MASS \citepads{2016ApJS..224...36K,2014ApJ...787..126L,2016ApJ...817..112S,
2003AJ....126.2421C,2001A&A...380..590P,2000AJ....120.1541M},
LP, NLTT, LHS \citepads{1979nlcs.book.....L,1979lccs.book.....L},
SCR \citepads{2007AJ....133.2898F},
SIPS \citepads{2007A&A...468..163D}).
There are also three common parallax and proper motion (CPPM) companions 
(\object{LP 889-37B}, \object{BD+52 911B}, \object{GJ 283 B}) among them.
Their primaries are mid- and early-M dwarfs, and a WD, respectively. 
About two thirds of the new UCDs in Table~\ref{Tab_newUCDinS}
have previous trigonometric or photometric distance estimates,
which were in some cases based on measurements of their brighter 
companions. We list them as parallaxes rounded to integer values
and mention that their uncertainties were of the order of several
milli-arcseconds. Nevertheless, they agreed in most cases rather well
with the {\it Gaia} DR2 parallaxes. 
Only for \object{NLTT 14748} the previous parallax value 
was much larger than measured by {\it Gaia}, and, except for
\object{LP 816-10}, all objects with previous distance estimates
were already known to be within 20\,pc.
About half of the new UCDs in this group had already
spectral type information provided in SIMBAD (see references of
Table~\ref{Tab_newUCDinS}).

Among the 19 new UCDs in the second group without SIMBAD identification
(Table~\ref{Tab_newUCDnoS}) we found no distance nor spectral type estimates
in the literature, 
except for \object{Gaia DR2 3106548406384807680},
already discovered by \citetads{2018RNAAS...2..205M} 
at its {\it Gaia} DR2-based distance 
of about 13.9\,pc (marked in Fig.~\ref{Fig_CMDallnoknown}). 
The new UCDs in this second group 
and the seven new UCDs from the first group (Table~\ref{Tab_newUCDinS})
that had no previous distance nor spectral type estimates contain
all 12 L-type UCDs and all three M9.5 dwarfs found in this study.
We note that two mid-L-type new UCDs and two M9.5 dwarfs 
(as well as two M8-M8.5 dwarfs) are not shown in
Fig.~\ref{Fig_CMDallnoknown}, because they lack $G$$-$$RP$ colour
measurements and belong to subsample E.

%
%
\begin{table*}
\caption{New ultracool dwarfs in the {\it Gaia} DR2 20\,pc sample with SIMBAD identification}
\label{Tab_newUCDinS}
\centering
\fontsize{6.4pt}{0.90\baselineskip}\selectfont
\begin{tabular}{@{}l@{\hspace{2mm}}r@{\hspace{2mm}}r@{\hspace{2mm}}r@{\hspace{2mm}}r@{\hspace{2mm}}r@{\hspace{2mm}}r@{\hspace{2mm}}r@{\hspace{2mm}}r@{\hspace{2mm}}r@{\hspace{2mm}}r@{\hspace{2mm}}c@{\hspace{2mm}}c@{\hspace{2mm}}l@{\hspace{2mm}}r@{\hspace{2mm}}l@{}}     
\hline\hline
SIMBAD & RA        & DE        & $Plx$ & $pm$RA  & $pm$DE  & $M_G$ & $G$$-$$RP$ & $M_J$ & $M_H$ & $M_{K_s}$ & Lit.$Plx$ &SIMBAD    & phot.  & $v_{tan}$ & Q \\
Name   &           &           &       &         &         &       &        &       &       &           & tri/pho   & SpT (Ref) & SpT    &           &   \\
       & (degrees) & (degrees) & (mas) & (mas/yr)& (mas/yr)& (mag) & (mag)  & (mag) & (mag) & (mag)     & (mas)&           &        & (km/s)    &   \\
\hline
\object{LSPM J0025+5422} &  6.314290&$+$54.380298& 59.45$\pm$0.12& $-$342.74$\pm$0.14& $-$376.19$\pm$0.13&14.61&1.52&10.65&10.09& 9.69& 68 T (11) & -&M7.5$\pm$1.0\tablefootmark{a}&   41 &A\\
\object{2MASS J01581572+1807128} & 29.565101&$+$18.120624& 50.84$\pm$0.18&  $-$88.20$\pm$0.23&  $+$86.78$\pm$0.19&14.19&1.45&10.50& 9.93& 9.60&     -     &-&M6.5$\pm$2.0\tablefootmark{a}&   12 &A\\
\object{LP 889-37B} & 62.231268&$-$31.482721& 55.78$\pm$0.15&  $-$19.34$\pm$0.20& $-$217.40$\pm$0.44&14.77&   -&    -&    -&    -& 64 T (12)\tablefootmark{b} &-&M8.0$\pm$1.0&                    19 &E\\
\object{LSPM J0449+5138} & 72.269109&$+$51.643140& 54.39$\pm$0.14&  $-$36.68$\pm$0.17& $-$373.36$\pm$0.13&14.61&1.51&10.50& 9.78& 9.42& 61 T (11) &-&M7.0$\pm$1.0\tablefootmark{a}&   33 &A\\
\object{BD+52 911B}\tablefootmark{c} & 75.856375&$+$53.122082& 72.52$\pm$0.18&$+$1286.92$\pm$0.31&$-$1570.93$\pm$0.30&15.03&1.63&    -&    -&    -& 71 T (13)\tablefootmark{b} &-&M8.5$\pm$0.5&                   133 &D\\
\object{NLTT 14748}\tablefootmark{e} & 79.223914&$+$56.670247& 68.83$\pm$0.11& $+$140.54$\pm$0.15& $-$408.28$\pm$0.14&14.23&1.47&10.35& 9.76& 9.38& 98 T (11) &-&M6.5$\pm$1.5\tablefootmark{a}&   30 &A\\
\object{LP 718-5} & 83.840233& $-$9.519472& 52.18$\pm$0.13& $+$343.89$\pm$0.24& $-$230.67$\pm$0.20&14.39&1.51&10.41& 9.77& 9.42& 52 P (1) &M6.5Ve (1)  &M7.0$\pm$1.0\tablefootmark{a}&   38 &A\\
\object{LSPM J0540+6417} & 85.094863&$+$64.283877& 57.48$\pm$0.20& $-$323.87$\pm$0.18& $-$105.53$\pm$0.20&16.10&   -&11.60&10.93&10.54&     -     &-&M9.5$\pm$1.0&                    28 &E\\
\object{2MASS J06320891-1009269} & 98.037603&$-$10.158467& 48.68$\pm$0.82&  $+$98.03$\pm$1.37& $-$210.37$\pm$1.65&18.65&1.77&13.43&12.58&11.94&     -     &-&L5.0$\pm$0.5&                    23 &B\\
\object{2MASS J06431389+1631428} &100.808283&$+$16.528867& 49.82$\pm$0.16& $+$112.94$\pm$0.28&  $+$62.30$\pm$0.23&14.23&1.48&10.34& 9.72& 9.35&     -     &-&M6.5$\pm$1.0\tablefootmark{a}&   12 &A\\
\object{GJ 283 B} &115.085915&$-$17.415063&109.05$\pm$0.08&$+$1152.39$\pm$0.11& $-$536.55$\pm$0.12&14.17&1.47&10.34& 9.82& 9.48& 107 T (14) &M6.5Ve (2)  &M6.5$\pm$1.5\tablefootmark{a}&   55 &A\\
\object{2MASS J08334323-5336417} &128.428118&$-$53.609780& 50.41$\pm$0.09& $-$256.01$\pm$0.20& $+$388.87$\pm$0.18&14.64&1.53&10.53& 9.86& 9.49&     -     &-&M7.5$\pm$0.5\tablefootmark{a}&   44 &A\\
\object{SCR J0838-5855} &129.508533&$-$58.934257& 90.04$\pm$0.22&  $-$61.82$\pm$0.55& $-$317.83$\pm$0.41&14.18&1.54&10.08& 9.48& 9.04& 93 T (15) &M6e (3)  &M6.5$\pm$0.5\tablefootmark{a}&   17 &C\\
\object{LP 788-1} &142.841347&$-$17.295745& 72.18$\pm$0.27& $-$367.20$\pm$0.35& $-$152.13$\pm$0.34&14.29&1.48&10.37& 9.76& 9.36& 71 T (15) &M6.5 (4)  &M6.5$\pm$1.0\tablefootmark{a}&   26 &C\\
\object{LP 848-50} &160.672533&$-$24.267132& 96.15$\pm$0.47&  $+$40.79$\pm$0.77& $+$156.58$\pm$0.69&14.01&1.45&10.19& 9.59& 9.25& 94 T (15) &-&M6.0$\pm$1.5\tablefootmark{a}&    8 &C\\
\object{2MASS J11155037-6731332} &168.965807&$-$67.526850& 57.53$\pm$0.09& $+$534.69$\pm$0.17& $-$227.09$\pm$0.18&15.21&1.57&10.88&10.20& 9.76&     -     &-&M8.5$\pm$0.5\tablefootmark{a}&   48 &A\\
\object{LP 793-34}\tablefootmark{d} &176.398174&$-$20.351067& 51.76$\pm$0.10& $+$148.82$\pm$0.18&  $+$68.73$\pm$0.11&14.10&1.47&10.30& 9.69& 9.35& 50 T (13)\tablefootmark{b} &M5e (5)  &M6.5$\pm$1.5\tablefootmark{a}&   15 &A\\
\object{LP 675-7}\tablefootmark{e} &185.965979& $-$8.979938& 52.26$\pm$0.12& $-$231.61$\pm$0.25& $-$275.78$\pm$0.14&14.29&1.48&10.45& 9.90& 9.55& 52 P (16) &M6.5Ve (1)  &M6.5$\pm$1.5\tablefootmark{a}&   33 &A\\
\object{LP 911-56} &206.690122&$-$31.823075& 73.23$\pm$0.16& $-$343.36$\pm$0.23& $+$158.62$\pm$0.26&14.18&1.48&10.30& 9.76& 9.36& 75 P (17) &M6 (5)  &M6.5$\pm$1.0\tablefootmark{a}&   24 &A\\
\object{NLTT 37185} &215.763247&$+$51.775978& 57.70$\pm$0.10& $+$362.66$\pm$0.16& $+$193.42$\pm$0.14&14.69&1.52&10.69&10.11& 9.76& 54 T (18) &-&M7.5$\pm$1.5\tablefootmark{a}&   34 &A\\
\object{2MASS J14241870-3514325} &216.077896&$-$35.242757& 52.39$\pm$0.24&   $-$8.05$\pm$0.45&  $-$76.49$\pm$0.32&14.41&1.50&10.44& 9.79& 9.40& 54 P (19) &M6.5 (6)  &M7.0$\pm$1.0\tablefootmark{a}&    7 &C\\
\object{LHS 2930} &217.650558&$+$59.724252&100.29$\pm$0.08& $-$808.01$\pm$0.14& $+$157.25$\pm$0.13&14.87&1.53&10.80&10.15& 9.79& 104 T (20) &M6.5V (7)  &M7.5$\pm$1.0\tablefootmark{a}&   39 &A\\
\object{LP 624-54} &243.604961& $-$2.848627& 68.70$\pm$0.13&   $-$7.22$\pm$0.23& $+$367.83$\pm$0.13&14.43&1.50&10.49& 9.87& 9.46& 68 P (5) &M5.0V (8)  &M7.0$\pm$1.0\tablefootmark{a}&   25 &A\\
\object{PM J17189-4131} &259.726030&$-$41.531726& 88.79$\pm$0.18& $-$220.05$\pm$0.17& $-$911.54$\pm$0.15&14.32&1.49&10.35& 9.74& 9.36&     -     &M6 (3)  &M6.5$\pm$1.0\tablefootmark{a}&   50 &A\\
\object{SIPS J1848-8214}\tablefootmark{e} &282.211515&$-$82.246211& 56.93$\pm$0.10&  $-$50.53$\pm$0.14& $-$272.51$\pm$0.20&14.38&1.51&10.26& 9.70& 9.28& 57 T (15) &-&M6.5$\pm$1.0\tablefootmark{a}&   23 &A\\
\object{2MASS J20021341-5425558}\tablefootmark{e} &300.556473&$-$54.433768& 55.30$\pm$0.11&  $+$61.13$\pm$0.15& $-$365.00$\pm$0.13&14.23&1.47&10.33& 9.75& 9.36& 58 P (17) &M6e (5)  &M6.5$\pm$1.5\tablefootmark{a}&   32 &A\\
\object{LP 695-372} &310.689258& $-$5.004822& 59.64$\pm$0.10& $+$241.33$\pm$0.16& $+$117.30$\pm$0.11&14.08&1.46&10.33& 9.75& 9.42& 51 P (17) &M6.5 (9)  &M6.5$\pm$1.5\tablefootmark{a}&   21 &A\\
\object{LSPM J2049+3336} &312.363261&$+$33.612383& 59.73$\pm$0.09& $-$207.68$\pm$0.13& $-$412.07$\pm$0.13&14.28&1.48&10.38& 9.75& 9.42& 68 T (11) &-&M6.5$\pm$1.5\tablefootmark{a}&   37 &A\\
\object{LP 816-10} &312.471081&$-$17.269443& 53.39$\pm$0.10& $+$301.91$\pm$0.17&  $-$98.57$\pm$0.11&14.41&1.49&10.45& 9.85& 9.44& 45 P (17) &M6 (10)  &M7.0$\pm$1.0\tablefootmark{a}&   28 &A\\
\object{2MASS J21010483+0307047} &315.274368& $+$3.117811& 54.34$\pm$0.16&$+$1023.18$\pm$0.27&  $-$55.15$\pm$0.26&14.56&1.54&10.38& 9.64& 9.24& 56 T (21) &-&M7.5$\pm$1.0\tablefootmark{a}&   89 &A\\
\object{SIPS J2151-4017} &327.883041&$-$40.290912& 56.11$\pm$0.15& $+$402.56$\pm$0.17& $-$265.50$\pm$0.18&14.16&1.49&10.19& 9.51& 9.16&     -     &-&M6.5$\pm$1.0\tablefootmark{a}&   41 &A\\
\hline
\end{tabular}
\tablefoot{\fontsize{6.4pt}{0.90\baselineskip}\selectfont
{\it Gaia} DR2 coordinates are for (J2000, epoch 2015.5) and were
rounded to 0.000001 degrees, parallaxes and their errors to
0.01\,mas, proper motions and their errors to 0.01\,mas/yr. 
Absolute magnitudes and colours derived from {\it Gaia} DR2 and
2MASS data were rounded to 0.01\,mag, tangential velocities
to 1\,km/s.
Parallaxes from the literature 
(T = trigonometric, P = photometric,
all rounded to 1\,mas) and spectral types from SIMBAD
are also given, if available. 
Photometric spectral
types were estimated using the relationships between
spectral types and absolute magnitudes and colours
given by \citetads{2018A&A...619L...8R} using
all available data, except for 
\tablefoottext{a}{($G-J$ not used)}.
Spectral type uncertainties represent
standard deviations of the values determined from different
relations for a given object.
\tablefoottext{b}{parallax of primary},
\tablefoottext{c}{companion discovered by \citetads{2015MNRAS.449.2618W},}
\tablefoottext{d}{found as common proper motion companion of
\object{LP 793-33}=\object{Hip 57361} by \citetads{2004ApJS..150..455G}.}
\tablefoottext{e}{with non-zero YMG membership probability given in
Table~\ref{Tab_YMG}.}
Q indicates
astrometric and photometric quality (subsamples A-E,
see Sect.~\ref{SubSect_q}).}
\tablebib{\fontsize{7pt}{0.90\baselineskip}\selectfont
(1) \citetads{2003AJ....126.3007R},
(2) \citetads{2015AJ....149..106D},
(3) \citetads{2017A&A...600A..19P},
(4) \citetads{2013A&A...556A..15R},
(5) \citetads{2006A&A...446..515P},
(6) \citetads{2003AJ....126.2421C},
(7) \citetads{1991AJ....101..662B},
(8) \citetads{2012AJ....144...99D},
(9) \citetads{2004AJ....128..463R},
(10) \citetads{2001A&A...380..590P},
(11) \citetads{2014ApJ...784..156D}, 
(12) \citetads{2017AJ....154..151B},
(13) \citetads{1997A&A...323L..49P},
(14) \citetads{2005AJ....130..337C}, 
(15) \citetads{2017AJ....153...14W},
(16) \citetads{2005A&A...442..211S},
(17) \citetads{2015AJ....149....5W},
(18) \citetads{2016AJ....151..160F},
(19) \citetads{2003AJ....126.2421C},
(20) \citetads{1992AJ....103..638M},
(21) \citetads{2017AJ....154..191J}.
}
\end{table*}

%
%
\begin{table*}
\caption{New ultracool dwarfs in the {\it Gaia} DR2 20\,pc sample without SIMBAD identification}
\label{Tab_newUCDnoS}
\centering
\fontsize{6.4pt}{0.90\baselineskip}\selectfont
\begin{tabular}{@{}l@{\hspace{3mm}}r@{\hspace{2mm}}r@{\hspace{3mm}}r@{\hspace{2mm}}r@{\hspace{2mm}}r@{\hspace{3mm}}r@{\hspace{2mm}}r@{\hspace{2mm}}r@{\hspace{2mm}}r@{\hspace{2mm}}r@{\hspace{2mm}}l@{\hspace{2mm}}r@{\hspace{2mm}}l@{}}     
\hline\hline
{\it Gaia} DR2 & RA        & DE        & $Plx$ & $pm$RA  & $pm$DE  & $M_G$ & $G$$-$$RP$ & $M_J$ & $M_H$ & $M_{K_s}$ & phot. & $v_{tan}$ & Q \\
Identifier     &           &           &       &         &         &       &        &       &       &           & SpT   &           &   \\
               & (degrees) & (degrees) & (mas) & (mas/yr)& (mas/yr)& (mag) & (mag)  & (mag) & (mag) & (mag)     &       & (km/s)    &   \\
\hline
\object{Gaia DR2 3195979005694112768}& 63.195222& $-$7.571338& 59.69$\pm$0.34& $+$407.80$\pm$0.46& $-$429.58$\pm$0.29&17.41&1.64&    -&10.92&10.80&L1.0$\pm$1.0&                    47  &B\\
\object{Gaia DR2 3432218798435750016}\tablefootmark{e}& 95.821390&$+$26.524936& 48.56$\pm$0.71&   $-$7.68$\pm$1.81& $-$131.87$\pm$1.97&18.18&1.71&13.10&11.97&11.30&L4.5$\pm$2.5&                    13  &D\\
\object{Gaia DR2 2923887295576778496}&100.105867&$-$23.872644& 49.89$\pm$0.73& $+$191.54$\pm$0.87& $-$134.23$\pm$1.52&18.88&1.75&13.56&12.58&11.93&L5.5$\pm$1.0&                    22  &B\\
\object{Gaia DR2 3106548406384807680}\tablefootmark{e,h}&100.903030& $-$2.387939& 71.92$\pm$1.38&  $+$28.28$\pm$2.34& $-$221.22$\pm$2.48&19.96&1.93&14.77&13.66&12.91&L9.0$\pm$2.0&                    15  &D\\
\object{Gaia DR2 5535283658436274944}\tablefootmark{e}&113.536854&$-$43.506480& 48.25$\pm$0.84&  $-$39.14$\pm$1.81& $+$151.66$\pm$2.01&19.14&1.86&13.98&13.00&12.43&L6.5$\pm$1.0&                    15  &B\\
\object{Gaia DR2 5424690587034891264}\tablefootmark{e,f}&143.353445&$-$43.893269& 57.03$\pm$0.10& $-$190.58$\pm$0.18& $+$162.25$\pm$0.17&14.20&1.48&10.32& 9.66& 9.30&M6.5$\pm$1.0\tablefootmark{a}&   21  &A\\
\object{Gaia DR2 5424690587034982144}\tablefootmark{e,f}&143.358169&$-$43.892756& 56.91$\pm$0.10& $-$196.75$\pm$0.17& $+$165.15$\pm$0.16&14.14&1.48&10.26& 9.61& 9.21&M6.5$\pm$1.5\tablefootmark{a}&   21  &A\\
\object{Gaia DR2 5432670704985289088}&145.874266&$-$38.566282& 77.72$\pm$0.09&  $-$95.11$\pm$0.13& $-$161.64$\pm$0.15&14.19&1.46&10.35& 9.82& 9.48&M6.5$\pm$1.5\tablefootmark{a}&   11  &A\\
\object{Gaia DR2 5229173470875537664}&157.802139&$-$73.385923& 54.09$\pm$0.13&  $+$39.63$\pm$0.23&  $-$59.37$\pm$0.21&15.94&   -&11.47&10.90&10.42&M9.5$\pm$1.0&                     6  &E\\
\object{Gaia DR2 3460907947316392704}&179.858244&$-$36.581076& 50.45$\pm$0.27&  $+$78.84$\pm$0.31& $-$140.94$\pm$0.20&15.87&1.68&11.07&10.35& 9.82&L0.0$\pm$1.0&                    15  &D\\
\object{Gaia DR2 1477762323724646272}\tablefootmark{g}&214.942956&$+$31.619184& 52.65$\pm$0.10& $+$104.21$\pm$0.17&  $-$21.29$\pm$0.17&14.07&1.46&10.30& 9.70& 9.36&M6.5$\pm$1.5\tablefootmark{a}&   10  &A\\
\object{Gaia DR2 6031367499416648192}\tablefootmark{e}&252.101014&$-$29.219946& 50.00$\pm$0.76&  $-$62.37$\pm$1.47& $-$115.01$\pm$0.69&18.50&1.70&13.30&12.29&11.63&L4.5$\pm$1.5&                    12  &B\\
\object{Gaia DR2 4169232338868860160}&262.863382& $-$7.997935& 54.17$\pm$0.17&  $+$40.76$\pm$0.28& $-$227.52$\pm$0.26&15.20&   -&10.93&10.25& 9.82&M8.5$\pm$0.5\tablefootmark{a}&   20  &E\\
\object{Gaia DR2 4171383636445972096}&271.776456& $-$6.430673& 53.40$\pm$0.47& $+$124.01$\pm$0.48& $-$206.12$\pm$0.38&16.26&1.63&11.63&10.98&10.49&L0.0$\pm$0.5&                    21  &B\\
\object{Gaia DR2 4159791176135290752}&277.826216& $-$7.540994& 54.00$\pm$0.40& $+$136.96$\pm$0.72& $-$154.84$\pm$0.67&15.89&1.71&11.16&10.36& 9.86&L0.0$\pm$1.5&                    18  &C\\
\object{Gaia DR2 4283084190940885888}&279.045710& $+$3.257012& 54.59$\pm$0.85& $+$256.58$\pm$1.37& $+$175.69$\pm$1.58&18.54&   -&13.48&12.56&11.92&L5.0$\pm$1.0&                    27  &E\\
\object{Gaia DR2 4206320171755704320}&286.529838& $-$5.251398& 49.12$\pm$0.61& $+$223.82$\pm$0.71&  $-$45.20$\pm$0.59&15.88&1.64&11.37&10.72&10.27&M9.5$\pm$0.5&                    22  &D\\
\object{Gaia DR2 6759481141756109056}&290.376628&$-$29.262960& 66.67$\pm$0.22& $+$512.18$\pm$0.30& $+$250.76$\pm$0.26&14.68&1.53&10.57& 9.97& 9.56&M7.5$\pm$1.0\tablefootmark{a}&   41  &A\\
\object{Gaia DR2 2034222547248988032}&298.988643&$+$32.255061& 59.34$\pm$0.93& $-$381.90$\pm$1.54& $-$560.21$\pm$1.49&19.14&   -&14.14&13.20&12.60&L6.5$\pm$2.0&                    54  &E\\
\hline
\end{tabular}
\tablefoot{\fontsize{6.4pt}{0.90\baselineskip}\selectfont
\tablefoottext{f}{\object{Gaia DR2 5424690587034891264} and \object{Gaia DR2 5424690587034982144} form a CPPM pair (angular separation $\approx$12.4\,arcsec),
as also found by \citetads{2019JDSO...15...21K}}
\tablefoottext{g}{CPPM with the known M4 dwarf \object{UCAC3 244-106602} (angular separation $\approx$9\,arcsec),}
\tablefoottext{h}{already discovered in {\it Gaia} DR2
and classified as L8 
by \citetads{2018RNAAS...2..205M}.}
Other notes and footmarks as given for Table~\ref{Tab_newUCDinS}.
}
\end{table*}

\section{Photometric classification of new UCDs}
\label{Sect_photclass}

\begin{figure}
   \centering
   \includegraphics[width=\hsize]{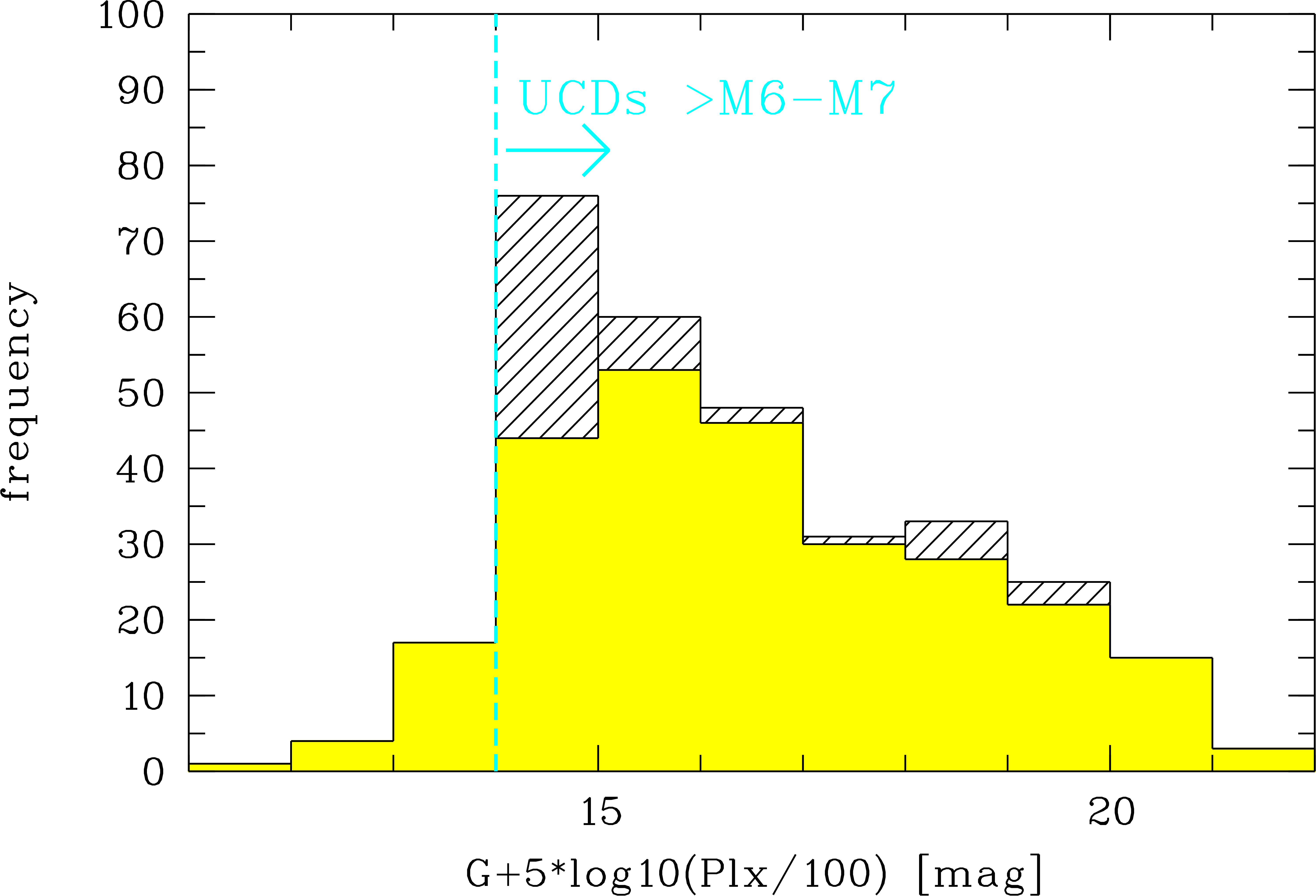}
      \caption{Distribution of {\it Gaia} DR2 absolute
              magnitudes $M_G$ of previously known 263 UCDs
              (filled yellow histogram) and of 50 newly found UCDs
              (added hashed black histogram) in the {\it Gaia} DR2
              20\,pc sample. The dashed line indicates
              the lower limit of $M_G>14$\,mag used in the search
              for new nearby UCDs.
              }
      \label{Fig_Gabshis}
   \end{figure}

From Fig.\ref{Fig_Gabshis} it becomes clear that our new
UCD additions in the {\it Gaia} DR2 20\,pc sample 
considerably filled the lowest absolute magnitude bin
14\,mag$<M_G<$15\,mag. This was expected, as there
was no strict absolute magnitude limit in the UCD selections 
of \citetads{2019ApJ...883..205B,2019MNRAS.485.4423S},
and \citetads{2018A&A...619L...8R}. Instead, their
samples were selected based on spectral types, 
where \citetads{2019MNRAS.485.4423S} tended to mainly include
L and T dwarfs, whereas the other two studies aimed at a
complete census of M-type UCDs, too. Our search was also
quite successful in the range 15\,mag$<M_G<$20\,mag,
roughly corresponding to spectral types between M8 and L8. 

For the photometric classification of our new UCDs we first 
cross-identified our list with the 2MASS and found high-quality
NIR photometry ($JHK_s$ quality flags ''AAA'') for almost all of 
our targets. Only in a few cases, because of bright companions
or merged background objects, no $JHK_s$ magnitudes
(\object{LP 889-37B}, \object{BD+52 911B}) were available, or 
the $J$ magnitude was missing (\object{Gaia DR2 3195979005694112768}).
In Tables~\ref{Tab_newUCDinS} and \ref{Tab_newUCDnoS}, we list all the 
corresponding optical and NIR absolute magnitudes and
the optical ($G-RP$) colour indices, if available. 

To estimate 
mean spectral types, we applied 
four spectral type-absolute magnitude ($M_G$, $M_J$, $M_H$, and $M_{K_s}$)
and three spectral type-colour ($G-RP$, $G-J$, and $J-K_s$)
relations as given in \citetads[][their Table 1]{2018A&A...619L...8R}.
For 27 out of 31 entries in Table\ref{Tab_newUCDinS}, 
and five out of 19 entries in Table~\ref{Tab_newUCDnoS},
the relation between spectral type and $G-J$ colour given 
by \citetads{2018A&A...619L...8R}
was not used as it was only valid for M9-T6 dwarfs. For the
objects with partly lacking photometric measurements, the spectral
types were estimated based on fewer data and are therefore less reliable.
The given uncertainties are the standard deviations 
of the values determined from the different relations
for a given object. The mean values
and their uncertainties were rounded to 0.5 spectral types. Typical
uncertainties fall in the range from $\pm$1.0 to $\pm$1.5 spectral types.
Only 16\% (8 of 50 objects in Tables~\ref{Tab_newUCDinS} and
\ref{Tab_newUCDnoS}) have more precise $\pm$0.5 spectral types.
On the other hand, for only 8\%, including one M6.5 dwarf in 
Table~\ref{Tab_newUCDinS} and three L4.5-L9 dwarfs in 
Table~\ref{Tab_newUCDnoS}, the uncertainties were larger
(from $\pm$2.0 to $\pm$2.5 spectral types).
For the M8 dwarf \object{LP 889-37B}, 
lacking $G$$-$$RP$ colour and $JHK_s$ 
photometry, we adopted an uncertainty of $\pm$1.0 spectral types.

We mentioned that the estimated spectral types for M-type
UCDs appeared systematically earlier from spectral type-colour relations 
than from spectral type-absolute magnitude relations. Nevertheless,
we preferred to use the mean spectral types derived
from all available data, rather than to restrict our photometric
classification to spectral type-absolute magnitude relations only.
Previously determined spectral types of nearly half of the objects in
Table~\ref{Tab_newUCDinS}, all of relatively early-type (M5-M6.5) UCDs, 
are in good agreement with our photometric spectral type estimates.
Except for our M7$\pm$1.0 dwarf \object{LP 624-54}, formerly
classified as M5 by \citetads{2012AJ....144...99D}, their spectral types
agree within our uncertainties.
This supported our decision to use all spectral type relations
listed in Table~1 of \citetads{2018A&A...619L...8R} in our
photometric classification.

Among the new UCDs with SIMBAD identification 
(Table~\ref{Tab_newUCDinS}) there are only few late-M (M8-M9.5)
dwarfs and only one L dwarf (with a spectral type of L5). 
However, Table~\ref{Tab_newUCDnoS}, listing
the new UCDs without previous SIMBAD identification, contains
systematically later types with nearly 60\% of them being
classified as L dwarfs, including seven $\gtrsim$L5 dwarfs.

With the UCD definition by an absolute magnitude cut
applied in this study, it is clear that all our new UCD
candidates will need spectroscopic follow-up. In particular those
with photometrically estimated spectral types at or below the
classical UCD spectral type dividing line M6.5/M7 may
later turn out not to be classical UCDs. However, we note that 
three of the UCD candidates of \citetads{2018A&A...619L...8R} with 
photometrically estimated spectral types of $\ge$M7 and even 15
of the spectroscopically classified by \citetads{2019ApJ...883..205B} 
classical UCDs with $\ge$M7 
fall below our absolute magnitude cut at $M_G=14$\,mag.
Because of the already mentioned relatively large spread in 
absolute magnitudes observed for known M7 dwarfs 
(Sect.\ref{SubSect_ucdwd}), we found it justified to include 
a relatively large number of M6.5 dwarfs and one M6 dwarf,
as classified by us photometrically, in our new UCD sample
presented in Tables~\ref{Tab_newUCDinS} and \ref{Tab_newUCDnoS}.

\section{Sky distribution and kinematics of new UCDs}
\label{Sect_kinspat}

\begin{figure}
   \centering
   \includegraphics[width=\hsize]{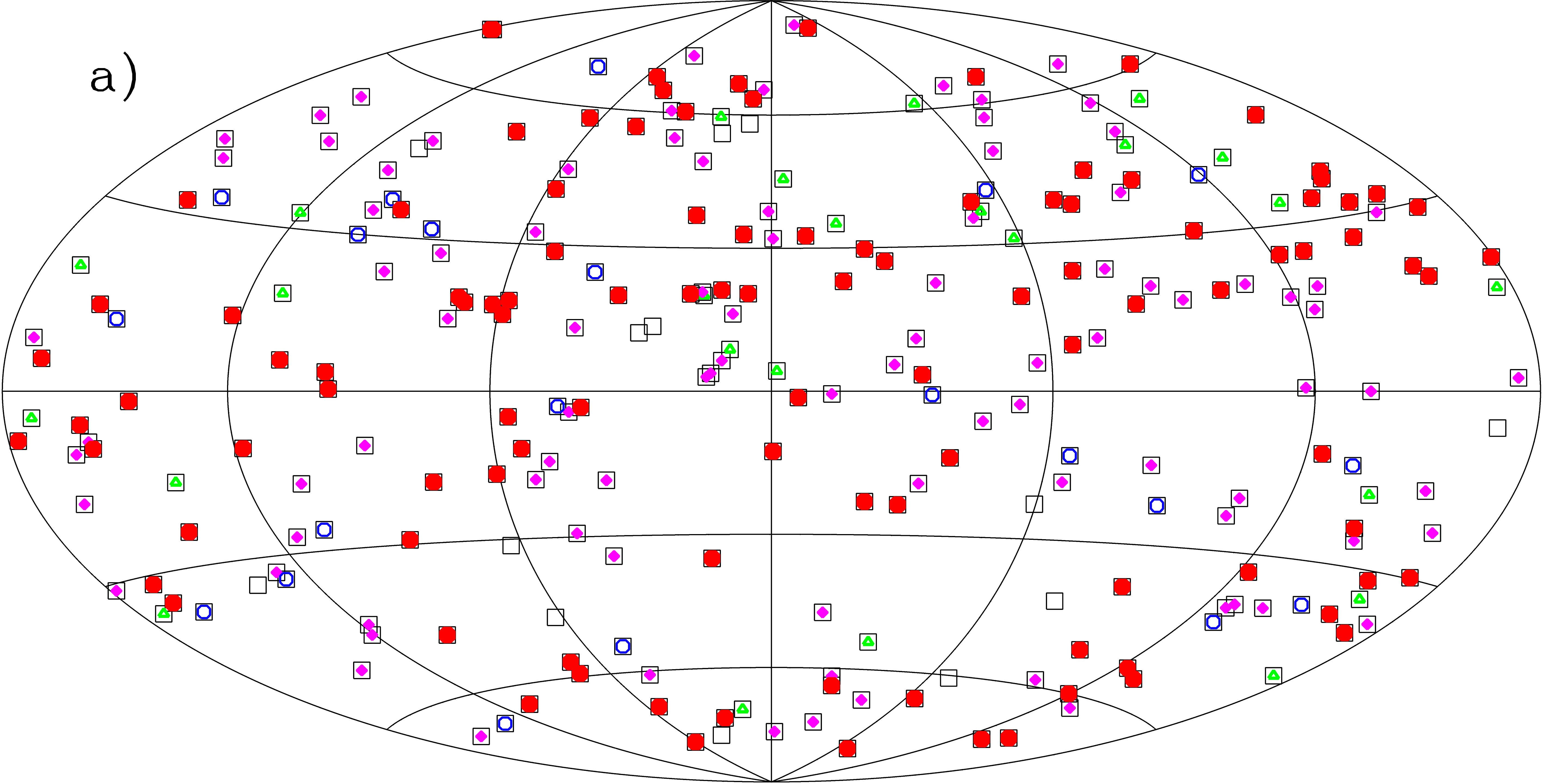}
   \includegraphics[width=\hsize]{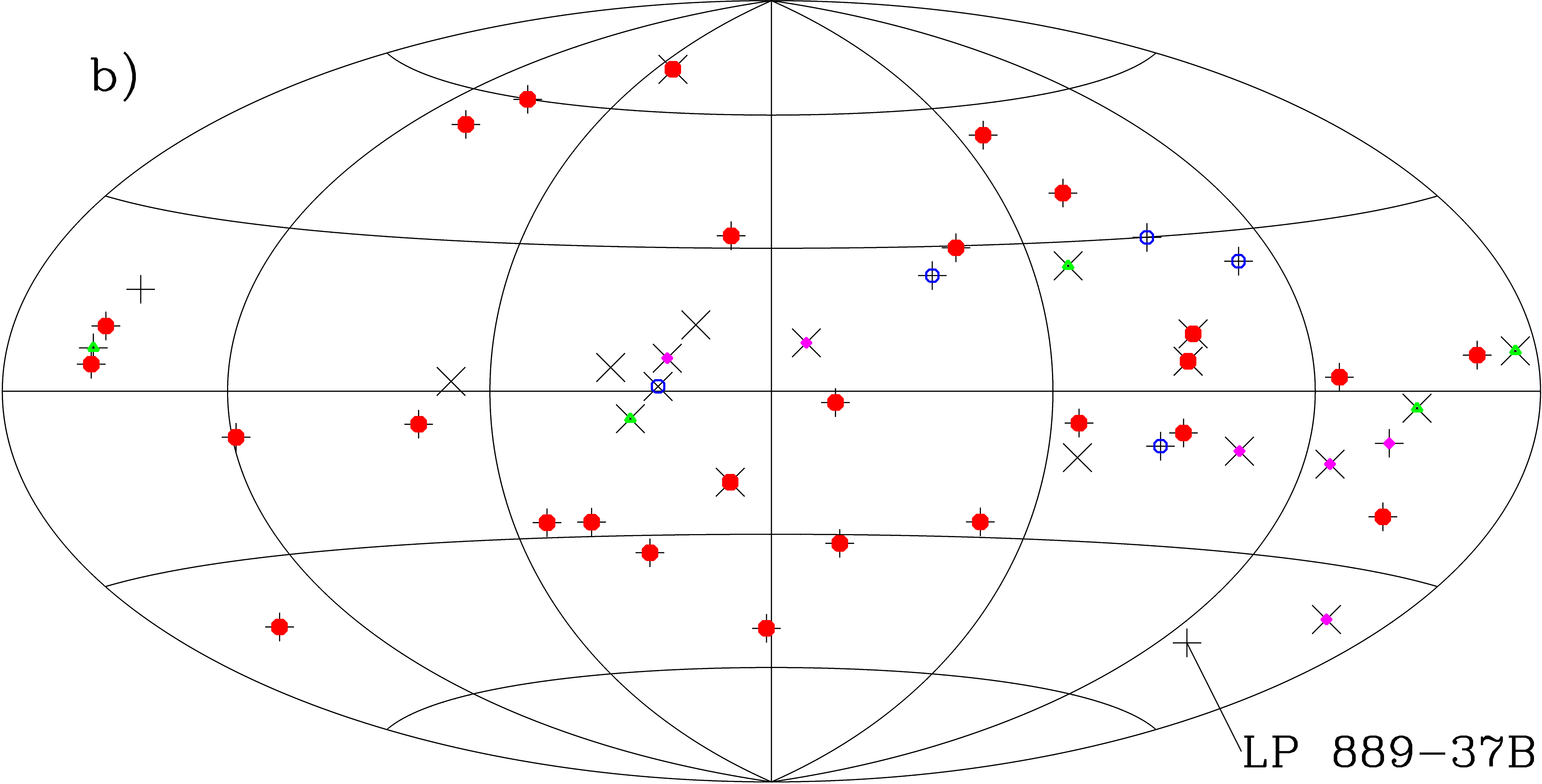}
      \caption{Sky distributions in Galactic coordinates 
              ($l,b$) of previously known UCDs (panel a)
              and new UCDs (panel b) in the {\it Gaia} DR2
              20\,pc sample. The map is centred 
              on $l=0\degr, b=0\degr$, with $l$ rising to the left, 
              and the Galactic north pole $b=90\degr$ at the top.
              The coloured filled and open symbols are as used
              in Fig.~\ref{Fig_CMDall} for the subsamples A, B, C, 
              and D, corresponding to different
              astrometric and photometric quality. Overplotted
              open squares in panel a) mark all 263 known UCDs
              (including 14 without $BP$$-$$RP$ colour 
              belonging to subsample E, hence no coloured symbols).
              Overplotted pluses (identified in SIMBAD) and crosses 
              (not found in SIMBAD) in panel b)
              mark all 50 new UCDs (including 6 of subsample E).
              The new UCD labelled
              in panel b) is a close companion of a previously known
              HPM star resolved by {\it Gaia} (see text).
              }
      \label{Fig_skyUCD}
   \end{figure}

Because of the general difficulty to identify UCDs in crowded
fields, we expected some incompleteness of previous searches, e.g. 
along the Galactic plane. In fact, the
sky distributions of previously known and new UCDs are
remarkably different (Fig.~\ref{Fig_skyUCD}). The distribution 
of the 263 previously known UCDs reveals some lack of objects in 
the southern Galactic hemisphere. The north-south asymmetry 
accounts 
to 148:115 UCDs. There is also a deficit of objects 
within $\pm15\degr$ from the Galactic equator - only 21\% fall in 
that Galactic latitude zone, whereas with a uniform sky distribution 
about 26\% would be expected.
On the other hand, the new UCDs exhibit a north-south 
ratio of 27:23, and are mainly found close to the Galactic equator 
- 54\% of them are in the Galactic equatorial zone with $|b|<15\degr$.
This is more than two times larger than expected with a uniform
distribution. A comparison of the extended pole regions with
$|b|>45\degr$ also leads to different results for the old and new 
UCD samples. Out
of 263 previously known UCDs, 76 fall in these regions. This
corresponds exactly to the about 29\% expected with a uniform 
distribution. But only 6 of the 50 new UCDs can be counted
in the extended pole regions, which corresponds to only 12\%.
Note that the concentration of new UCDs towards the Galactic plane
is even stronger for those without SIMBAD identification
(crosses in Fig.~\ref{Fig_skyUCD}).

\begin{figure}
   \centering
   \includegraphics[width=\hsize]{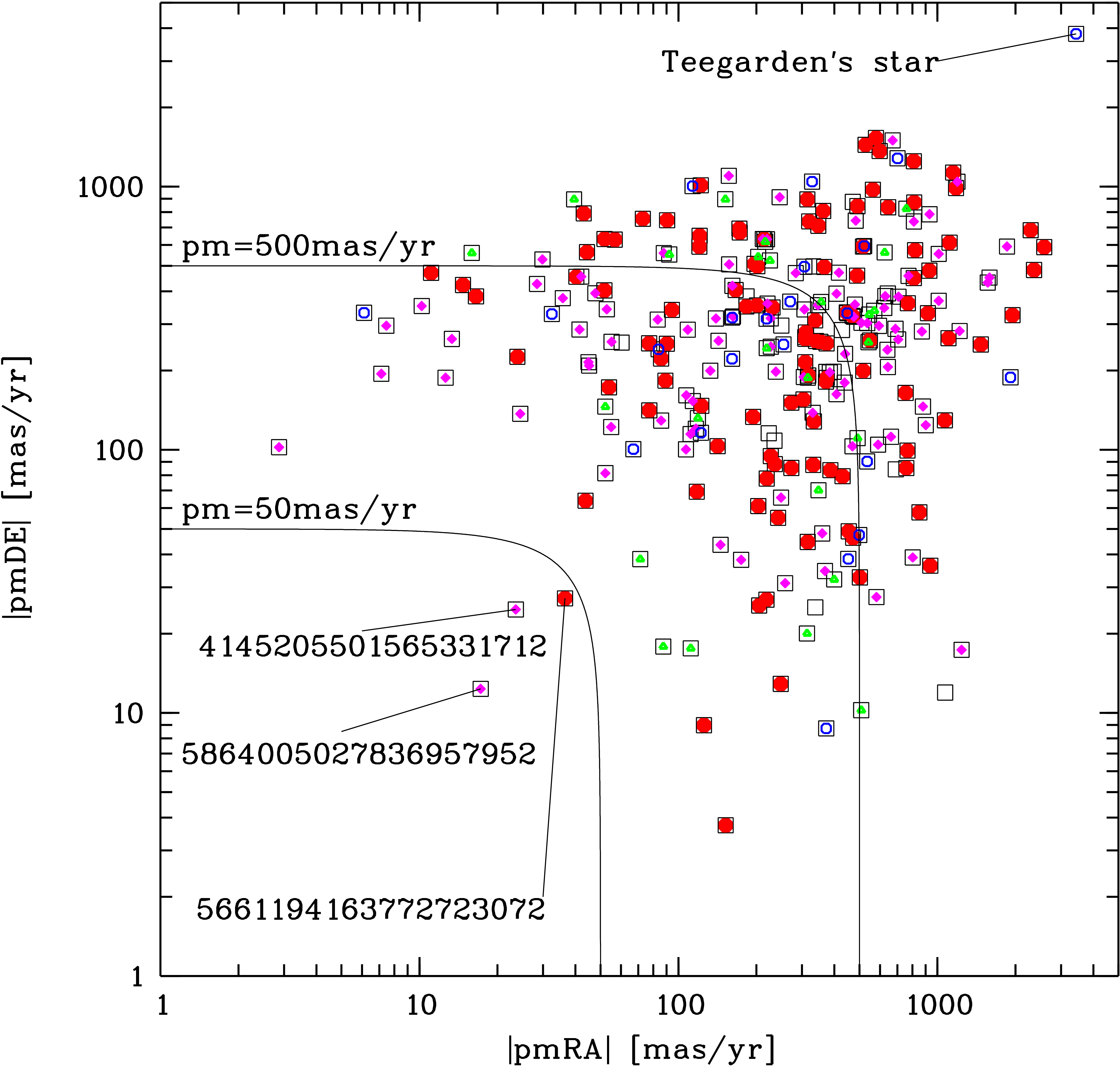}
   \includegraphics[width=\hsize]{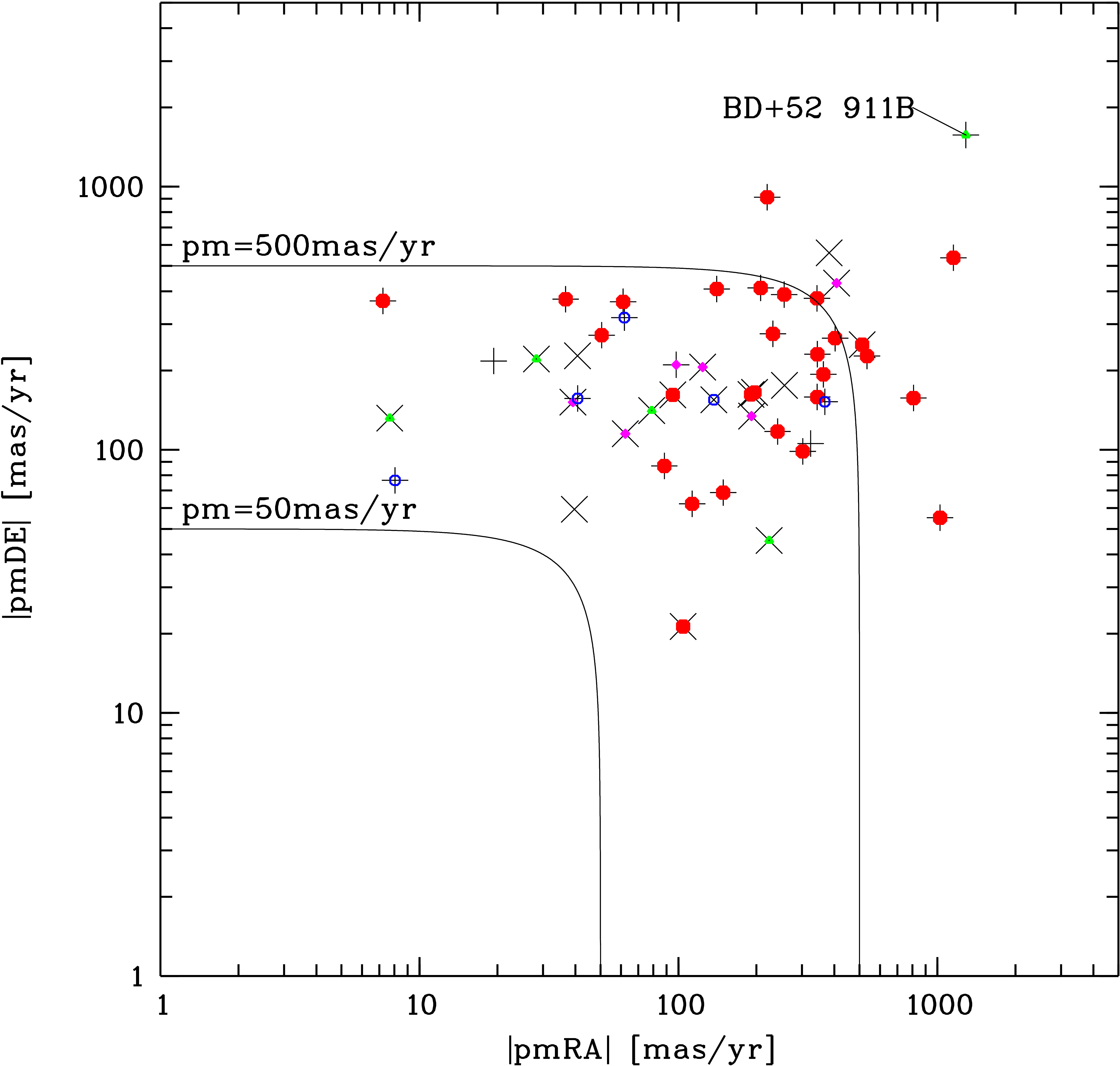}
      \caption{Absolute proper motion components $|pmRA|$
              and $|pmDE|$
              of all previously known UCDs (top panel)
              and new UCDs (bottom panel) in the {\it Gaia} DR2
              20\,pc sample in logarithmic scale.
              Total proper motions of 
              50\,mas/yr and 500\,mas/yr 
              are indicated by solid lines.
              All symbols are as used in Fig.~\ref{Fig_skyUCD}.
              Objects with extremely large and small proper motions
              are marked by their name or {\it Gaia} DR2 numbers.
              }
      \label{Fig_pmUCD}
   \end{figure}

The proper motions of the new UCDs are typically not as large as
those of the previously known ones (Fig.~\ref{Fig_pmUCD}). Whereas
nearly half (48\%) of the known UCDs had total proper motions 
larger than 500\,mas/yr, only 10 out of 50 (20\%) new UCDs were
found with such large proper motions. 
Even if we account for the fact that all newly found UCDs in 
the {\it Gaia} DR2 20\,pc sample are at distances larger than 9\,pc,
we still notice relatively small proper motions.
If we select only the 234 known UCDs 
at a distance $>$9\,pc, then 43\% of them (101 objects)
have $>$500\,mas/yr proper motion. This fraction
is still more than twice as large as that of the new ones.
A trend towards smaller proper motions of new neighbours was expected, 
since 
HPM surveys \citepads[e.g.][]{1979nlcs.book.....L,
2005AJ....129.1483L,
2005A&A...442..211S,
2007A&A...468..163D,
2014ApJ...787..126L,
2016ApJS..224...36K}
played an important role in the
search for nearby UCDs in the pre-{\it Gaia} era.

However, among the new UCDs we did also not find such extremely 
small proper motions $<$50\,mas/yr as observed for three known UCDs. 
These three known UCDs as well as one with the largest proper motion 
are marked in the top panel of Fig.~\ref{Fig_pmUCD}, respectively.
The latter is \object{Teegarden's star} \citepads{2003ApJ...589L..51T},
listed in \citetads{2019ApJ...883..205B} as an M7 dwarf and confirmed by
the {\it Gaia} DR2 parallax of 261\,mas as the second nearest known UCD
(the nearest in that list and of all 263 known UCDs in our {\it Gaia} DR2 
20\,pc sample is the well-known M6.5 dwarf \object{GJ 1111} 
with a parallax of 279\,mas). Interestingly, all three known UCDs with 
the smallest proper motions, 
\object{Gaia DR2 4145205501565331712},
\object{Gaia DR2 5661194163772723072}, and
\object{Gaia DR2 5864005027836957952}, were found as UCD candidates
in {\it Gaia} DR2
by \citetads{2018A&A...619L...8R} and photometrically classified as
L1.5, M8.0, and M9.5 dwarfs, respectively. One of them,
\object{Gaia DR2 5661194163772723072} was also mentioned in
a nearby M dwarfs multiplicity survey 
by \citetads{2019AJ....157..216W}. When we look again at the top
right panel of Fig.~\ref{Fig_apm} and compare it with the top
panel of Fig.~\ref{Fig_pmUCD}, then we note that there are
good Solar neighbour candidates with small proper motion
and earlier spectral types.
However, these
possibly not yet identified early- to mid-M dwarf neighbours
were not subject of this investigation.

As in Fig.~\ref{Fig_CMDallnoknown}, the new UCDs are plotted in
Figs.\ref{Fig_skyUCD} and \ref{Fig_pmUCD} with plus signs, when they
were already identified in SIMBAD (Table~\ref{Tab_newUCDinS}), and with 
crosses, when they were not (Table~\ref{Tab_newUCDnoS}).
In both Figs.~\ref{Fig_skyUCD} and \ref{Fig_pmUCD} the different
astrometric and photometric quality of the known and new UCDs
can be seen from the underplotted coloured symbols. There are 
no obvious systematic effects with quality, 
neither in the distributions
on the sky (panel a) in Fig.\ref{Fig_skyUCD})
nor in the absolute proper motion diagrams of the known
UCDs (top panel in Fig.~\ref{Fig_pmUCD}). In the bottom
panel of Fig.~\ref{Fig_pmUCD} one can see that the new UCDs
with SIMBAD identification (pluses) have on average larger
total proper motions than those without SIMBAD identification
(crosses). There is another clear trend visible in the sky
distribution of the new UCDs (panel b) in Fig.\ref{Fig_skyUCD}),
if we exclude the Galactic plane by cutting at $b>30\degr$ and
$b<-15\degr$. Outside of such an asymmetric Galactic plane zone,
two times broader in the north than in south,
there is a lack of open coloured
symbols (corresponding to subsamples C and D) and only one 
exceptional case (see below) of a missing underplotted coloured 
symbol in the south (corresponding to subsample E). This means, 
outside of this Galactic plane region, almost all new
UCD neighbours appear to be astrometrically well-behaved.

The one exception, \object{LP 889-37B} marked in panel b) of
Fig.~\ref{Fig_skyUCD}, is the fainter component of a close binary
(separation 1.2\,arcsec) resolved in {\it Gaia} DR2 as a CPPM
pair (with a mean parallax of $\approx$55.5\,mas), as already
mentioned by \citetads{2019JDSO...15...21K}. The primary, 
\object{LP 889-37A}, appears 3\,mag brighter ($G\approx$13.0\,mag) and
astrometrically well-behaved ($RUWE=1.18$), whereas \object{LP 889-37B}
is lacking the $BP$ and $RP$ measurements and can also be classified
as a bad astrometry object ($RUWE=2.81$). The previously unresolved
HPM star 
\object{LP 889-37} was included in the catalogue
of bright M dwarfs \citepads{2011AJ....142..138L} and classified 
by \citetads{2017AJ....154..151B} as a nearby M4 dwarf with a much
larger trigonometric parallax ($\approx$64\,mas) than measured by
{\it Gaia} DR2. The parallax measurement of \citetads{2017AJ....154..151B}
was probably affected by the unresolved companion.

In Tables~\ref{Tab_newUCDinS} and \ref{Tab_newUCDnoS} we also
list the tangential velocities of the new UCDs 
computed based on {\it Gaia} DR2 data. For comparison, we note that
\citetads{2019MNRAS.485.5573T} have
studied the different Galactic populations, thin disk, thick disk,
and halo, of WDs within 100\,pc and compared {\it Gaia} DR2
results with simulated data. They showed that the lower limits of
the observed tangential velocity of thick disk and halo stars are at
90\,km/s and 200\,km/s, respectively. However, there is a large
overlap of the tangential velocity distributions of thin disk, thick
disk and halo stars. According to Fig.~10 of \citetads{2019MNRAS.485.5573T},
the tangential velocities of thick disk stars may be as low as
50\,km/s, while those of thin disk stars may reach up to about
100\,km/s. The highest tangential velocities of thick disk stars
are around 160\,km/s, but halo stars may have minimum tangential
velocities of about 120\,km/s.
 
As a consequence of the already noted systematic differencees in the
total proper motions (bottom
panel of Fig.~\ref{Fig_pmUCD}), the 
tangential velocities of the new UCDs without SIMBAD identification
(Table~\ref{Tab_newUCDnoS}) are systematically lower than
those of the new UCDs with SIMBAD identification
(Table~\ref{Tab_newUCDinS}). Whereas all new UCDs without SIMBAD 
identification have $v_{tan}$$\lesssim$50\,km/s and probably belong 
to the thin disk population, among those with SIMBAD identification
there are two with much larger $v_{tan}$. One of them, 
\object{2MASS J21010483+0307047} has $v_{tan}$$\approx$90\,km/s, which lies
just at the above mentioned limit between thin and thick disk stars. This
HPM star 
was discovered by \citetads{2000AJ....120.1541M},
mentioned as a binary by \citetads{2017AJ....154..191J}, and 
reported as an object with clear astrometric evidence of orbital motion 
\citepads[][see their Fig.~12]{2019AJ....157..216W} but not resolved
in {\it Gaia} DR2. 
The new UCD with the highest tangential velocity is
\object{BD+52 911B} 
\citepads{2015MNRAS.449.2618W}
with $v_{tan}\approx$130\,km/s
(also marked as the new UCD with the
most extreme HPM 
in Fig.~\ref{Fig_pmUCD}). 
The primary of this resolved {\it Gaia} DR2 CPPM companion 
(separation 5.6\,arcsec), \object{BD+52 911}, is a long-known high 
tangential velocity star \citepads[e.g.][]{1984AJ.....89..720L}
included in the catalogues of bright M dwarfs \citepads{2011AJ....142..138L}
and of {\it Gaia} DR2 radial velocity standard stars, and targetted by
many spectroscopic surveys like HADES \citepads{2017A&A...598A..26P}
and CARMENES \citepads{2018A&A...614A..76J}.
From comparison with the population study of \citetads{2019MNRAS.485.5573T},
we conclude that this system belongs most probably to the Galactic thick disk.
Using the {\it Gaia} DR2 data for \object{BD+52 911}, including its 
radial velocity of 65.8$\pm$0.2\,km/s, we computed Galactic space 
velocity \citepads{1987AJ.....93..864J} components 
of $(U,V,W)=(-113.6\pm0.2,-93.6\pm0.1,+14.5\pm0.1)$\,km/s.  
According to Fig.~9 (bottom panel) of \citetads{2016MNRAS.463.2453D} 
showing 3$\sigma$ $UV$ 
velocity ellipsoids for different Galactic populations 
taken from \citetads{2000AJ....119.2843C}, the $UV$ values fall outside
the Galactic thin disk and within the thick disk boundaries, confirming our 
former conclusion based on tangential velocities alone.

\section{Membership in young moving groups}
\label{Sect_YMG}

To estimate the possible YMG membership of our 50 new UCDs we used 
the Bayesian Analysis for Nearby Young AssociatioNs (BANYAN) of 
\citetads{2018ApJ...856...23G}. Their tool provided 
at http://www.exoplanetes.umontreal.ca/banyan/
allows for kinematic membership determination in 29 known YMGs.
As input parameters we used the coordinates, proper motions and
parallaxes, but no radial velocities. Those UCDs that had non-zero
membership probabilities in at least one of the YMGs are listed
in Table~\ref{Tab_YMG} and marked by a footnote in Tables~\ref{Tab_newUCDinS}
and \ref{Tab_newUCDnoS}.

\begin{table}
\caption{YMG membership probabilities of new UCDs}              
\label{Tab_YMG}      
\fontsize{7.0pt}{0.90\baselineskip}\selectfont
\centering                                      
\begin{tabular}{@{}l@{\hspace{2mm}}r@{\hspace{3mm}}r@{\hspace{4mm}}r@{\hspace{4mm}}r@{\hspace{4mm}}r@{}}     
\hline\hline                        
Object name                           & ABDMG & CARN & COL  & ARG  & Field \\    
                                      & (\%)  & (\%) & (\%) & (\%) &  (\%) \\
\hline                                   
\object{NLTT 14748}                   &  2.4  & 95.4 &  0.1 &  0.1 &  2.0  \\
\object{LP 675-7}                     & 99.6  &  -   &   -  &   -  &  0.4  \\        
\object{SIPS J1848-8214}              &  2.3  &  -   &   -  &   -  & 97.7  \\
\object{2MASS J20021341-5425558}      & 64.2  &  -   &   -  &   -  & 35.8  \\
\object{Gaia DR2 3432218798435750016} &  -    &  0.2 &   -  & 99.0 &  0.8  \\
\object{Gaia DR2 3106548406384807680} &  -    &  -   &  2.4 &   -  & 97.6  \\
\object{Gaia DR2 5535283658436274944} &  -    &  0.3 &   -  &  5.5 & 94.2  \\
\object{Gaia DR2 5424690587034891264} &  -    &  -   &   -  & 40.8 & 59.2  \\
\object{Gaia DR2 5424690587034982144} &  -    &  -   &   -  & 71.4 & 28.5  \\
\object{Gaia DR2 6031367499416648192} &  -    &  -   &   -  & 71.0 & 29.0  \\
\hline                                             
\end{tabular}
\tablefoot{\fontsize{7.0pt}{0.90\baselineskip}\selectfont
Only non-zero YMG membership probabilities and the corresponding field
membership probabilities are listed as computed using BANYAN 
\citepads{2018ApJ...856...23G}.}
\end{table}

All of the possible YMG members listed in Table\ref{Tab_YMG}
have membership probabilities ranging between 0.1\% and 99.6\%,
and belong to one of the following four YMGs:
AB Doradus \citepads[ABDMG;][]{2004ApJ...613L..65Z},
Carina Near \citepads[CARN;][]{2006ApJ...649L.115Z},
Columba \citepads[COL;][]{2008hsf2.book..757T}, and
Argus \citepads[ARG;][]{2000MNRAS.317..289M}.
The ABDMG has, according to \citetads{2005ApJ...628L..69L} 
and \citetads{2013ApJ...766....6B}, an age of $\approx$125\,Myr,
comparable with that of the Pleiades. A relatively old age of 
about 200\,Myr \citepads{2006ApJ...649L.115Z} is still assumed for 
CARN, but this YMG has only been sparsely 
investigated since its discovery.
\citetads{2015MNRAS.454..593B} determined a young age of about 42\,Myr
for COL and failed to assign an unambiguous age to ARG, because of
a contaminated membership list. However, \citetads{2019ApJ...870...27Z}
recently determined a young age of about 40-50\,Myr for ARG, too.

Concerning COL, there are only very low membership probabilities
determined for two of our UCDs. On the other hand, three of the UCDs listed 
in Table~\ref{Tab_YMG} have very high membership probabilities, between 
95\% and 100\%, in one of the other three YMGs. Of these three, neither the 
two with SIMBAD identifications (\object{NLTT 14748} in CARN
and \object{LP 675-7} in ABDMG) nor the one without 
(\object{Gaia DR2 3432218798435750016} in ARG) were previously
known as YMG members, although one of them, \object{LP 675-7} was included 
in the input catalogue of \citetads{2015ApJ...798...73G}, who searched for
UCD members in YMGs. Another three entries of Table~\ref{Tab_YMG} show
still relatively large membership probabilities between about 64\% and 71\%.
Again, we found that the one of them already identified in SIMBAD
\object{2MASS J20021341-5425558} in ABDGM) was included in the input list 
of \citetads{2015ApJ...798...73G}. 
The other two,
\object{Gaia DR2 5424690587034982144} and
\object{Gaia DR2 6031367499416648192}, join the highly-probable ARG member
\object{Gaia DR2 3432218798435750016}. The CPPM companion of
\object{Gaia DR2 5424690587034982144}, \object{Gaia DR2 5424690587034891264}
has a lower 41\% membership probability in the same YMG. The many
ARG members among our new UCDs, including two photometrically
classified L4.5 dwarfs (\object{Gaia DR2 3432218798435750016} and
\object{Gaia DR2 6031367499416648192}) are remarkable, as they
may represent relatively low-mass young brown dwarfs.

\section{Discussion and outlook}
\label{Sect_outlook}

Some of the nearest known members of the 20\,pc sample were not included in 
{\it Gaia} DR2 or have only photometry but no parallax and proper motion
measurements, because they are close binaries that could not be
well described by the 5-parameter astrometric solution. This
concerned all of the nearest known UCD binaries including L and T dwarfs:
\object{WISE J1049-5319AB}
\citepads[L7.5+T0.5 at 2.0\,pc;][]{2013ApJ...767L...1L,2013ApJ...772..129B,
2018A&A...618A.111L}, 
\object{eps Indi B}a+Bb 
\citepads[T1+T6 at 3.6\,pc;][]{2003A&A...398L..29S,2004A&A...413.1029M,
2018ApJ...865...28D},
and \object{SCR J1845-6357}A+B
\citepads[M8.5+T6 at 3.9\,pc][]{2004AJ....128..437H,2006ApJ...641L.141B}.
The first one, \object{WISE J1049-5319AB}, was resolved in
{\it Gaia} DR2 as two nearly equally bright components 
($G$$\approx$16.95\,mag) with an angular separation of about 0.9\,arcsec,
both without proper motion and parallax. The second, \object{eps Indi B}a+Bb,
was discovered by \citetads{2004A&A...413.1029M} with an angular separation 
of about 0.7\,arcsec in August 2003 and should have appeared with
similar separations during the {\it Gaia} observational epochs about 12
years later, since the orbital period was found to be about 11.4\,yr
by \citetads{2018ApJ...865...28D}. However, \object{eps Indi B}a+Bb
was not resolved in {\it Gaia} DR2 and listed with photometry only
($G$$\approx$18.6\,mag). The third nearest UCD binary,
\object{SCR J1845-6357}A+B, was found with an angular separation of
about 1.2\,arcsec in 2005 \citepads{2006ApJ...641L.141B}. The {\it Gaia} DR2
provided full astrometry of the M8.5 primary (with $G$$\approx$14.0\,mag) but
no other object within an angular separation of 3.5\,arcsec.
A very similar system of a late-M dwarf with a close mid-T dwarf companion
is \object{WISE J072003.20-084651.2}
\citepads[M9.5+T5.5 at 6.8\,pc;][]{2014A&A...561A.113S,2015AJ....149..104B,
2019AJ....158..174D}. At the expected position of this system, 
{\it Gaia} DR2 contained only an unresolved source ($G$$\approx$15.3\,mag) 
lacking the parallax and proper motion.
This binary 
had, according to \citetads{2019AJ....158..174D},
separations of the order of only 0.2\,arcsec at the
{\it Gaia} DR2 epochs.

Some known WDs within 20\,pc, including four within 10\,pc,
are missing from {\it Gaia} DR2, or are listed without parallax and 
proper motion measurements. \citetads{2018MNRAS.480.3942H} attempted 
to provide explanations for this.
Interestingly,
\object{Sirius B} had a measured parallax (Fig.~\ref{Fig_CMDallknown}), 
whereas \object{Procyon B} was not included in {\it Gaia} DR2
due to saturation of its very bright primary 
(separation $\approx$4\,arcsec). 
Another very nearby WD companion of a bright star, 
\object{40 Eri B} (separation $\approx$8\,arcsec), was listed
without parallax and proper motion. The WD \object{Wolf 489},
which was missing although not known to be affected by a companion 
or merged background star, was noted by \citetads{2018MNRAS.480.3942H}
as having an extremely large total proper 
motion of $\approx$4\,arcsec/yr. This may have led to matching 
problems in the {\it Gaia} DR2 data reduction.
The fourth WD within 10\,pc listed without parallax and proper motion
is \object{G 99-47}, for which \citetads{2018MNRAS.480.3942H} do
not give a note. As in case of \object{Wolf 489}, the
IRSA finder charts do not show any disturbing background object,
but \object{G 99-47} could be a close binary. 
\citetads{2017A&A...602A..16T} found that possibly
more than ten resolved double WDs were not yet recognised
in the {\it Gaia} DR2 20pc sample.

We speculate that some other nearby and apparently single objects 
may be also missing in {\it Gaia} DR2 because of their not yet
uncovered close binary nature. With the next data releases
of {\it Gaia}, the number of individual epoch measurements will
increase, but the astrometric solution will include more
parameters, taking into account possible orbital motion of close
companion(s). With the growing number of parameters one will
have to newly evaluate the astrometric and photometric quality,
probably based on a larger number of quality criteria than applied
in this work. Again, one can expect that some still problematic
nearby objects will be missing or be only partly supplied with the
astrometric parameters in {\it Gaia} DR3. For those objects
with at least given parallax and proper motion, we think that 
external proper motion checks and visual inspection of IRSA (or other)
finder charts will continue to play an important role. This will
again help us finding the real nearby stars among
probably still large numbers of false positives that we
expect in {\it Gaia} DR3, especially in crowded regions. 

The total number of UCDs increased to 313 objects.
We still observe a clear Galactic north-south assymetry of 175:138.
However, 83 of the 313 UCDs (26.5\%) fall within
$|b|<15\degr$. This is almost exactly what would be expected
with a uniform distribution. Therefore, as a result of our study,
this Galactic plane region 
can no longer be considered as incomplete with respect to UCDs.

We mentioned that our latest-type (L9)
new UCD, \object{Gaia DR2 3106548406384807680}
(= \object{WISE J064336.71-022315.4}; 
marked in Fig.~\ref{Fig_CMDallnoknown}),
was already discovered and spectroscopically classified
by \citetads{2018RNAAS...2..205M} as an L8 brown dwarf
with a distance of 13.9\,pc based on {\it Gaia} DR2 but
not yet included in SIMBAD. Our photometric estimate
of the spectral type is in reasonable good agreement
with the spectroscopic classification. The relatively small
tangential velocity (15\,km/s) but nearly ten times
larger radial velocity ($\approx$140\,km/s) measured
by \citetads{2018RNAAS...2..205M} led them conclude
that this object kinematically belongs to the
Galactic thick disk and passed close (only 1.4\,pc)
from the 
Sun 
just 100\,000 years ago. Currently,
only one even closer flyby is known in the recent
past, which is that of \object{WISE J072003.20-084651.2}
\citepads{2015ApJ...800L..17M}. This already mentioned
UCD binary with lacking {\it Gaia} DR2 astrometry
passed, according to the most recent results of
\citetads{2019AJ....158..174D}, at a distance of 0.33\,pc
from the Sun just 80\,000 years ago.

From their tangential velocities 
alone (Tables~\ref{Tab_newUCDinS} and \ref{Tab_newUCDnoS}),
almost all new UCDs can be considered thin disk members.
Only two of them have $v_{tan}\gtrsim$90\,km/s and are more
likely thick disk members. Especially those with the 
smallest tangential velocities ($v_{tan}\lesssim$10\,km/s)
may be interesting targets for
radial velocity measurements. Among them one could find
more such close encounters with the 
Sun,
where the total
radial velocity is much larger than the tangential
velocity. If such cases arise, they possibly represent
thick disk members crossing the solar neighbourhood.
As we were not able to confirm any new UCD neighbours
with extremely small proper motions, we may still
miss slowly moving thin disk members like
\object{GJ 710} that pass very close from the 
Sun.
\citetads{2018A&A...616A..37B} investigated the 
about seven million relatively bright stars with radial 
velocity measurements in {\it Gaia} DR2 and found that
\object{GJ 710} is the closest encounter (only 0.08\,pc
from the Sun.  With its exceptional (for stars within 20\,pc) 
zero proper motion and zero tangential velocity, 
and a radial velocity of only $-$14\,km/s, 
this is a clear representative of the Galactic thin disk. 

We found six most likely members in three YMGs (ABDGM, CARN, and ARG)
with membership probabilities
between 64\% and 100\%, one of which has a CPPM companion that has also
a relatively high membership probability of 41\% in the same YMG.
Five of these seven YMG members were photometrically classified by us 
as M6.5-M7 dwarfs, two of which had already comparable SIMBAD spectral 
types. There are also two L4.5 dwarfs that were found as 99\%
and 71\% members of ARG, the youngest of the three YMGs, where most
members were detected. These mid-L type UCDs may
represent very low-mass young brown dwarfs. Low- and medium-resolution
spectroscopy are needed to confirm the spectral types of all new
UCD YMG members and to search for spectral signatures of youth,
respectively.

Only two new UCDs were found within 10\,pc, \object{GJ 283 B} 
at about 9.2\,pc and \object{LHS 2930} at about 10.0\,pc.
Both were already known in SIMBAD as M6.5 
dwarfs 
and as members of the 10\,pc sample
(Table~\ref{Tab_newUCDinS}), and our photometric
spectral types were derived as M6.5 and M7.5, respectively. 
The next nearest new UCDs found (between 10\,pc and 15\,pc) 
are also predominantly of relatively early types (M6-M8.5)
with the exception of the late L dwarf 
\object{Gaia DR2 3106548406384807680} (L8 according to
\citetads{2018RNAAS...2..205M}, L9 from our photometric
classification) at about 14\,pc, close to the expected
{\it Gaia} distance limit predicted by \citetads{2017MNRAS.469..401S} 
for objects of this spectral type. The new UCDs found at distances
between about 15 and 20\,pc are more uniformly spread over spectral 
types (M6.5-L6.5). We have not found any new T dwarfs
within the expected {\it Gaia} distance limits (e.g. about 14\,pc for 
T0-T4 according to \citetads{2017MNRAS.469..401S}). This may be
because previous searches for nearby T dwarfs were more successful
and complete than for nearby L and late-M dwarfs.

Our UCDs were defined by absolute magnitude and need spectroscopic
confirmation, before some of them can be considered as classical UCDs. In 
Table~\ref{Tab_newUCDinS}, there are already 14 non-classical UCDs with
SIMBAD spectral types between M5 and M6.5, which we classified photometrically
as M6.5-M7.5 dwarfs. For another 12 (Table~\ref{Tab_newUCDinS}) and five
(Table~\ref{Tab_newUCDnoS}) only photometrically classified M6-M7.5 dwarfs
we can also expect that only part of them will be confirmed spectroscopically
as classical UCDs.
However, we think that at least the remaining five and 14 photometrically 
classified M8-L9 dwarfs in Tables~\ref{Tab_newUCDinS} and \ref{Tab_newUCDnoS} 
are in fact classical UCDs.

\begin{acknowledgements}
This work analyses results from the European Space Agency (ESA) space 
mission {\it Gaia}. {\it Gaia} data are being processed by the {\it Gaia} 
Data Processing and Analysis Consortium (DPAC). Funding for the DPAC is 
provided by national institutions, in particular the institutions 
participating in the {\it Gaia} MultiLateral Agreement (MLA). 
The {\it Gaia} mission website is https://www.cosmos.esa.int/gaia. 
The {\it Gaia} archive website is https://archives.esac.esa.int/gaia.
We have also extensively used SIMBAD and VizieR at the CDS/Strasbourg
and would like to thank the CDS staff for their valuable work. 
The detailed referee report was very helpful for imroving
the content and presentation of this paper.
\end{acknowledgements}

\bibliographystyle{aa}
\bibliography{aa37373_lit}

\end{document}